\documentclass[twocolumn, dvipsnames, twocolappendix]{aastex631}

\graphicspath{{./}{figures/}}

\usepackage{amsmath}	
\usepackage{graphicx}
\usepackage{xspace}

\newcommand\citecrisp{Thomas et al. (in prep)}
\newcommand\citepcrisp{(Thomas et al. in prep)}

\defcitealias{TimonGlobal2024}{T25}
\newcommand\tttf{\citetalias{TimonGlobal2024}\xspace}
\newcommand\tttfp{\citepalias{TimonGlobal2024}\xspace}

\newcommand{\mhdX}{MHD\xspace}
\newcommand{\advX}{CR-A\xspace}
\newcommand{\nlX}{CR-NL\xspace}
\newcommand{\inX}{CR-NL-IN\xspace}

\newcommand{\crmhd}{CRMHD\xspace}

\newcommand{\sech}{\mathrm{sech}}

\newcommand{\sign}{\mathrm{sign}}

\newcommand{\velsym}{v}

\usepackage{xpatch}
\usepackage{bm}
\usepackage{hyperref}

\hypersetup{
	colorlinks=true,
	breaklinks=true,
	citecolor=Blue,
	allcolors=Blue,
	frenchlinks=true
}

\makeatletter
\xpatchcmd\NAT@citex
{%
	\@citea\NAT@hyper@{%
		\NAT@nmfmt{\NAT@nm}%
		\hyper@natlinkbreak{\NAT@aysep\NAT@spacechar}{\@citeb\@extra@b@citeb}%
		\NAT@date
	}%
}
{%
	\@citea
	\NAT@nmfmt{\NAT@nm}%
	\NAT@aysep\NAT@spacechar
	\NAT@hyper@{\NAT@date}%
}
{}{}
\xpatchcmd\NAT@citex
{%
	\@citea\NAT@hyper@{%
		\NAT@nmfmt{\NAT@nm}%
		\hyper@natlinkbreak{\NAT@spacechar\NAT@@open\if*#1*\else#1\NAT@spacechar\fi}%
		{\@citeb\@extra@b@citeb}%
		\NAT@date
	}%
}
{
	\@citea
	\NAT@nmfmt{\NAT@nm}%
	\NAT@spacechar\NAT@@open\if*#1*\else#1\NAT@spacechar\fi
	\NAT@hyper@{\NAT@date}%
}
{}{}
\makeatother

\begin{document}

\title{Cosmic Ray-Driven Galactic Winds with Resolved ISM and Ion-Neutral Damping}

\correspondingauthor{Brandon Sike}
\email{bsike@umich.edu}

\author[0009-0008-1788-4355]{Brandon Sike}
\affiliation{Department of Astronomy, University of Michigan, Ann Arbor, MI 48109, USA}

\author[0000-0002-7443-8377]{Timon Thomas}
\affiliation{Leibniz Institute for Astrophysics, Potsdam (AIP), An der Sternwarte 16, D-14482 Potsdam, Germany}

\author[0009-0002-2669-9908]{Mateusz Ruszkowski}
\affiliation{Department of Astronomy, University of Michigan, Ann Arbor, MI 48109, USA}

\author[0000-0002-7275-3998]{Christoph Pfrommer}
\affiliation{Leibniz Institute for Astrophysics, Potsdam (AIP), An der Sternwarte 16, D-14482 Potsdam, Germany}

\author[0000-0002-2910-2276]{Matthias Weber}
\affiliation{Leibniz Institute for Astrophysics, Potsdam (AIP), An der Sternwarte 16, D-14482 Potsdam, Germany}

\begin{abstract}

Feedback processes in galaxies dictate their structure and evolution. Baryons can be cycled through stars, which inject energy into the interstellar medium (ISM) in supernova explosions, fueling multiphase galactic winds. Cosmic rays (CRs) accelerated at supernova remnants are an important component of feedback. CRs can effectively contribute to wind driving; however, their impact heavily depends on the assumed CR transport model. We run high-resolution ``tallbox'' simulations of a patch of a galactic disk using the moving mesh magnetohydrodynamics code \textsc{Arepo}, including varied CR implementations and the \textsc{Crisp} non-equilibrium thermochemistry model. We characterize the impact of CR feedback on star formation and multiphase outflows. While CR-driven winds are able to supply energy to a global-scale wind, a purely thermal wind loses most of its energy by the time it reaches $3~\mathrm{kpc}$ above the disk midplane. We further find that the adopted CR transport model significantly affects the steady-state of the wind. In the model with CR advection, streaming, diffusion, and nonlinear Landau damping, CRs provide very strong feedback. Additionally accounting for ion-neutral damping (IND) decouples CRs from the cold ISM, which reduces the impact of CRs on the star formation rate. Nevertheless, CRs in this most realistic model are able to accelerate warm gas and levitate cool gas in the wind but have little effect on cold gas and hot gas. This model displays moderate mass loading and significant CR energy loading, demonstrating that IND does not prevent CRs from providing effective feedback.

\end{abstract}


\section{Introduction}
\label{sec:introduction}

Outflows are a key component of feedback behavior in galaxies \citep{Veilleux2005, Zhang2018, Veilleux2020, Laha2021_AGN_ionized_outflows, ThompsonHeckman2024_winds}. Feedback processes shape the structure and evolution of galaxies by ejecting material, supplying pressure support, and halting accretion on to the interstellar medium \citep[ISM;][]{SomervilleDave2015_Galaxy_Formation_Review, NaabOstriker2017_theor_chall_gal_form}. Galactic inflows and outflows must necessarily be the channels through which galaxies interact directly with their circumgalactic medium \citep[CGM;][]{TumlinsonPeeplesWerk2017, Pakmor2020, vandeVoort2021, Faucher-GOh2023_CGM_theory, Heesen2023MagnetizedCGM, Heesen2024III}. Altogether, galactic feedback dictates how galaxies are able to self-regulate in response to their content and surroundings, as well as how baryons are processed throughout the universe.

Galaxies with $M_\mathrm{halo} \lesssim 10^{12}~\mathrm{M}_\odot$ are primarily regulated by stellar feedback \citep{McKee1977_sn_regulated_ISM, Moster2010_stellar_m_halo_m, MitchellSchaye2022_EAGLE_SHM}. Energy released from supernovae (SNe) can drive bipolar winds from the plane of disk galaxies \citep{Bordoloi2014_bipolar_outflows_mass_loading, Guo2023_bipolar_outflows_observed} and form superbubbles \citep[e.g.,][]{ClarkeOey2002_sn_overlap, Barnes2023, Huang2023_superbubble, Mayya2023} which are more dynamically effective than isolated SNe events. The resulting winds are characterized by their mass loading factors \citep[e.g.,][]{Bordoloi2014_bipolar_outflows_mass_loading, Xu2022_CLASSY_mass_loading}, molecular composition \citep{Bolatto2013_molecular_outflow, Fisher2024}, and their dynamics, whether manifesting as a complete outflow \citep{Irwin2024_radio_outflow} or undergoing recycling in a fountain-like process \citep{Rubin2022_fountains}. Any theoretical framework to describe these galactic outflows must be able to account for multiphase gas and outflow rates comparable to star formation rates.

Cosmic rays (CRs) are an intensely studied agent of galactic feedback, and have been a proposed mechanism to reconcile simulations with observations \citep[see reviews by][]{Zweibel2017, RuszkowskiPfrommer2023_CR_Review, Owen2023}. The bulk of CRs are accelerated to relativistic kinetic energies at SN shocks in the ISM \citep{BaadeZwicky1934_CRs_supernovae, BlandfordEichler1987} and subsequently populate the galaxy. CRs provide roughly equal pressure support in the ISM as compared to thermal, magnetic, and turbulent pressures \citep{BoularesCox1990}, suggesting a significant role of CRs in galactic dynamics. CRs have long been theorized to participate in driving galactic outflows \citep{Ipavich1975, Breitschwerdt1991, Everett2008_CR_driven_wind_milky_way, Socrates2008_CR_Eddington_Limit}, and are particularly favored due to their long cooling times \citep[e.g.,][]{Ensslin2007_cr_cooling, Ensslin2011_long_cr_cooling_times_ICM, Pfrommer2017_AREPO_CR} and strong coupling to the thermal gas \citep[e.g.,][]{Zweibel2013}. Additionally, hadronic $\gamma$-ray emissions in galactic haloes \citep[e.g.,][]{Salem2016, Pfrommer2017_simulating_gamma_ray_emission, Lopez2018, Karwin2019,Werhahn2021b,Werhahn2023_spectral_gamma} and a radio-FIR correlation \citep[nonthermal radio emission correlated with dust-reprocessed stellar light; e.g,][]{vanderKruit1971_FIR_Radio_I, Bell2003_eric_radio_FIR, Werhahn2021c,Pfrommer2022} imply that CRs are necessarily tied to galactic feedback. Radio observations of nearby edge-on galaxies show that star-forming galaxies can host extended extraplanar diffusive synchrotron haloes \citep{Irwin2024_radio_outflow} indicating the presence of an escape mechanism for galactic CRs from within galaxies into their surrounding inner CGM. The observed halo morphology and radio intensity profiles are well modelled by CR-driven galactic winds \citep{Stein2023, Chiu2024_radio}.

Results based on simulations including CR feedback generally share several conclusions: (i) CRs can drive outflows throughout the entire extent of the wind, rather than just injecting energy from the base \citep{Uhlig2012,SalemBryan2014, ThomasPfrommerPakmor2023_Wave_Dark, TimonGlobal2024}; (ii) CR-driven outflows are predicted to be cooler than outflows without CRs \citep[e.g.,][]{Peters2015_CRs_tallbox_xrays, Girichidis2018_cooler_and_smoother_crs, Farcy2022, Rathjen2023, TimonGlobal2024}; (iii) CRs can be the dominant pressure component in both the ISM and CGM \citep{Chan2022, Armillotta2022, ThomasPfrommerPakmor2023_Wave_Dark, TimonGlobal2024}. However, different assumptions about CR microphysics can lead to vastly different results \citep[e.g.,][]{Simpson2016, Ruszkowski2017, Hopkins2021_different_transport, Armillotta2021, HeintzZweibel2022_Analytic_CR_Eddington, DeFelippis2024_CR_CGM_diffusive}.

CRs behave as a relativistic fluid over galactic scales. The streaming (or gyro-resonant) instability \citep{KulsrudPearce1969, Skilling1975_Streaming, Shalaby2021, Shalaby2023, Lemmerz2024} is the process by which CRs scatter off of self-excited Alfvén waves and ``stream'' and/or diffuse along magnetic field lines. However, this process is time- and spatially-dependent due to effects including ion-neutral damping \citep[IND;][]{KulsrudPearce1969} of Alfvén waves. In practical applications, following this ``self-confined'' transport behavior requires two-moment transport models \citep{JiangOh2018, ThomasPfrommer2019_CRHD}, which co-evolve the CR flux and energy densities to simultaneously produce streaming and diffusive behavior. These models benefit from high spatial resolution both to capture tangled magnetic topology and to resolve CR interactions with cold cloud interfaces \citep{BustardZweibel2021}.

Clearly, there are many important aspects to modeling CR-driven galactic winds. Not only must the models be up to the task, but the resolution must be sufficiently high. For the presented work, we choose to neglect global-scale effects and instead prioritize well-resolved, ISM-scale physics in a realistic galactic wind. We achieve this by comparing simulations of a patch of a Milky Way-type galactic disk (a ``tallbox'' approach) using the moving-mesh MHD code \textsc{Arepo} with the \textsc{Crisp} (Cosmic Rays and InterStellar Physics) feedback framework \citepcrisp. We focus on the effects of progressive additions of CR physics so that we can fully characterize the impact of each component. This work is, in spirit, an extension of \citet{Farber2018}, who performed a tallbox simulation with two-zone diffusive CRs to emulate decoupling in cold gas; \citet{Rathjen2023}, who performed tallbox simulations with purely diffusive CRs and studied outflows; and \citet{Armillotta2021, Armillotta2022, Armillotta2024}, who considered CR transport by including IND and non-linear Landau damping (NLLD) via post-processing of MHD simulations, and performed short simulations that included the dynamical impact of CRs. This works also serves as a companion to \citet[][hereafter \tttf]{TimonGlobal2024}, who performed global simulations with the \textsc{Crisp} model.

The outline of this paper is as follows. In Section~\ref{sec:sim_setup}, we describe our numerical model including initial conditions, CR physics, and feedback model. Section~\ref{sec:structure_and_profiles} presents structural differences between the simulations. In Section~\ref{sec:star_formation_and_outflows}, we look in detail at the feedback processes and outflows. This includes star formation, outflows, and mass and energy loading factors. We complement the outflow analysis with a deeper investigation of the driving forces in Section~\ref{sec:outflow_driving}, specifically focusing on acceleration by CRs and the resulting velocity structure by phase. We compare our results with global simulations in Section~\ref{sec:discuss_t24}, and present our conclusions in Section~\ref{sec:summary}.

\section{Numerical Model and Simulation Setup}
\label{sec:sim_setup}

The simulations presented in this paper are performed using the moving-mesh code \textsc{Arepo} \citep{Springel2010_AREPO, Pakmor2016y_AREPO, Weinberger2020_AREPO_Public}, including self-gravity, the MHD module \citep{Pakmor2011_AREPO_MHD1, Pakmor2013_AREPO_MHD2}, the CR module \citep{Pfrommer2017_AREPO_CR}, and the extension for 2-moment Alfvén wave-regulated cosmic ray magnetohydrodynamics \citep[\crmhd;][]{ThomasPfrommer2019_CRHD, ThomasPfrommer2022_CRHD_AREPO_2, ThomasPfrommerPakmor2021_CRHD_AREPO}. An  appropriate summary of the \crmhd equations can be found in \citet{ThomasPfrommerPakmor2023_Wave_Dark} and a similar summary of the MHD equations in \citet{Pfrommer2017_AREPO_CR}. We use the \textsc{Crisp} framework to model various microphysical processes in the ISM \citepcrisp.

\subsection{\textsc{Crisp} Feedback Model}
\label{sec:crisp_ism_model}

The implementation of the \textsc{Crisp} model in this work is nearly identical to that in \citet{TimonGlobal2024} with several adjustments for our setup. A full outline of the numerical methods of the \textsc{Crisp} model is given in \citecrisp.  Here we give a brief overview of the \textsc{Crisp} modules used in this work.

\textsc{Crisp} models non-equilibrium thermochemical processes to follow gas energetics in a time-dependent, rigorous manner. The chemistry module tracks diatomic, neutral, and ionized hydrogen, neutral and two ionized states of helium, and neutral and singly-ionized states of carbon, oxygen, and silicon. The evolution of these species is calculated using reaction rates from radiative and collisional ionization, radiative and dielectric recombination, charge exchange reactions, dust-mediated processes, and CR ionization. These abundances are used for cooling, heating, and CR transport processes.

Interstellar gas cooling is governed by various radiative and collisional processes. Low-temperature fine-structure metal lines from ground state and singly ionized C, O, and Si cool the gas and are calculated directly from collision rates \citep{Abrahamsson2007, Grassi2014}. High-temperature metal line cooling is calculated by interpolating a table prepared with the \textsc{Chianti} code \citep{Dere1997}. Rotation-vibrational lines of H$_2$ \citep{Moseley2021}, Ly$\alpha$ cooling by H \citep{Cen1992}, and free-free emission at high temperatures \citep{Ziegler2018} contribute to cooling as well. Heating of the thermal gas in the ISM--aside from energetic feedback events--is most strongly mediated by photoelectric heating of dust grains and polycyclic aromatic hydrocarbons (PAHs) by absorption of FUV photons \citep{BakesTielens1994}. Energetic CR reactions \citep{Pfrommer2017_AREPO_CR} also contribute to ionization and heating of the gas. 

We differ from \citet{TimonGlobal2024} in our treatment of the FUV radiation field. The FUV radiation field in this work is height-attenuated based on absorption from an idealized disk profile, and takes the form of a numerical fit function. The details of this calculation are described in Appendix~\ref{sec:fuv_appendix}. Work on simulations of star-forming disks with radiative transfer shows that the FUV radiation field can become strongly spatially variable \citep{Linzer2024_tigress_rad}, but the impact of the FUV field on star formation is expected to be smaller than the impact of SNe and outflows \citep{Rathjen2024_SILCC_FUV}. We include radiation contribution by a meta-galactic UV background. The spectral energy density from \citet{Puchwein2019} at $z=0$ is used alongside cross-sections from \citet{Verner1996} and self-shielding according to \citet{Rahmati2013} to follow photoionization and heating of dilute gas. We neglect the pressure contribution from radiation as it is not expected to affect dynamics outside of a particular subset of environments \citep[e.g.,][]{LiBryanJOstriker2017_supernovae_outflows_tallbox}. Radiation pressure is also not expected to strongly alter statistical star-formation properties \citep{Rosdalh2015_SFR_Density_PDF}.

We model stars as collisionless particles and prescribe their formation with a standard Schmidt-type approach \citep{Schmidt1959_SF_efficiency_density, Kravtsov2003_schmidt_sfr}. Gas cells over a threshold density of $1000X^{-1}\mathrm{m}_{\mathrm{p}}~\mathrm{cm}^{-3}$ are assigned a star formation efficiency per free-fall time $\epsilon_{\mathrm{ff}}=100\%$, with $X=0.76$ as the hydrogen mass fraction for this density. This threshold is chosen to ensure that (i) stars are systematically formed in the densest regions of cold molecular clouds, and (ii) the clustering of newly born stars is consistent with observations \citep[e.g.,][]{Buck2019_SF_threshold, Keller2022_SF_Threshold}. This density threshold also has been shown to best reproduce observational star-formation rates and to prevent an excessive star forming episode at the beginning of the simulations \citep{Gatto2017_sink_density_threshold}. The chosen $\epsilon_{\mathrm{ff}}$ ensures rapid generation of star particles and prevents further gas collapse to unphysically high densities.

Feedback by star particles consists of type II SNe. These type II SNe inject mass, metals, and energy based on \textsc{Starburst99} calculations \citep{Leitherer1999_Starburst99_SNe}. Because we are resolving masses representative of individual stars, we take a modified approach as compared to the \textsc{Crisp} feedback model. Individual stars are instead chosen to become type II SNe based on a random draw to maintain a global SN rate consistent with the SFR. The probability is chosen to ensure an average of 1 SN (with energy $1.06\times10^{51}~\mathrm{erg}$) per $100~\mathrm{M}_\odot$ of stars formed. The production of star particles followed by their delayed demise as SNe constructs a multiphase, feedback-regulated ISM.

\subsection{\crmhd}
\label{sec:crhd_subsection}

Of particular importance to this work is the numerical implementation of CRs. We use the \crmhd theory of \citet{ThomasPfrommer2019_CRHD}, which treats CRs as a relativistic fluid and solves for CR proton flux and energy densities alongside their gyroresonant Alfvén waves. This theory follows a gray spectrum of CR protons with an average energy of $\sim$1 GeV, as these are the most abundant CRs in the solar neighborhood ISM \citep{Stone2019_voyager_crs} and are expected to be well-described by the self-confinement picture \citep{ThomasPfrommer2019_CRHD}. The implementation of the \crmhd module \citep{ThomasPfrommerPakmor2021_CRHD_AREPO, ThomasPfrommer2022_CRHD_AREPO_2} in \textsc{Arepo} is built upon the one-moment CR module from \citet{Pfrommer2017_AREPO_CR}. The two-moment \crmhd solver uses a subcycled reduced-speed-of-light approximation to evolve the CR quantities.

We follow the transport of self-confined CRs through their interactions with gyroresonant Alfvén waves \citep{KulsrudPearce1969, Shalaby2023}. The Alfvén waves themselves can be damped by various processes. The primary damping mechanism is NLLD, in which thermal ions dissipate the energy of Alfvén beat waves \citep{Miller1991}. A secondary damping mechanism is IND, sometimes called ion-neutral friction \citep{KulsrudPearce1969}. This damping mechanism occurs in partially neutral gas, where the thermal ions supporting an Alfvén wave transfer momentum to the neutral atoms through mutual scattering events, resulting in damping of Alfvén waves and, thus, in a reduction in CR self-confinement. Ion-neutral damping therefore results in a net ``decoupling'' of CRs from the ambient gas and significantly alters the dynamical effects of CRs. We use momentum-transfer cross-sections from \citet{PintoGalli2008} to calculate the damping rate due to collisions between species tracked by the \textsc{Crisp} model. These Alfvén wave energy losses are deposited into the gas by heating and by momentum transfer along magnetic field lines.

CRs lose energy by transferring it to the surrounding thermal gas through mechanisms such as collisions, scattering, and gas ionization. Coulomb losses occur as a result of scatterings by charged particles in the thermal plasma, causing a net transfer of energy to the gas. Hadronic losses are the result of inelastic interactions between CR protons and thermal nuclei, producing pions that subsequently decay. The energy released in this process can escape as gamma rays and secondary particles, while a fraction is thermalized. We follow the prescription for hadronic and Coulomb losses from \citet{Pfrommer2017_AREPO_CR} and distribute the thermalized energy to the gas in the \textsc{Crisp} chemistry solver.

The GeV CR protons of interest to this work are primarily accelerated at supernova remnant(SNR) shocks or stellar wind termination shocks in the ISM \citep{Grenier2015_CR_Nine_Lives}. Kinetically following CR acceleration at shocks is beyond the scope of this work. Instead, we inject CR energy as an additional $5\%$ of the total energy of the SN. This is an approximate angle-average of quasi-parallel and quasi-perpendicular CR acceleration at SNR shocks \citep{Pais2018,Winner2020_sn1006}.

\subsection{Initial Conditions and Numerical Setup}
\label{sec:ics_numerical_setup}

The bulk of the analysis for this work utilizes data outputs corresponding to simulation time greater than several tens of Myrs. At this time, the structure of the gas is strongly influenced by feedback activity. We present the initial setup for transparency, but note that the exact formulae describing the initial conditions do not reflect the actual state of the simulation at any analyzed time.

Each of the presented simulations start from an identical setup. We initialize a $1~\mathrm{kpc} \times 1~\mathrm{kpc} \times \pm 4~\mathrm{kpc}$ domain with periodic $x$- and $y$-boundary conditions and outflow-only (``diode'') $z$-boundary conditions. The diode boundaries do not allow mass to flow in from the boundary and prevent unphysical superheated inflows in tallbox simulations \citep{Hill2012}. For the initial gas profile, we begin with the form of an isothermal, self-gravitating disk in hydrostatic equilibrium \citep{Spitzer1942_sech2, Camm1950_sech2, SalemBryan2014, Ruszkowski2017}, and evaluate the profile at a fixed galactic radius \citep[e.g.,][]{Farber2018, Holguin2019, Holguin2024}. This gives a vertical gas mass density profile
\begin{equation} \label{eq:sech_profile}
    \rho(z) = \left[\left(\frac{\Sigma_{\mathrm{gas}}}{4 z_0} \right) \coth{\left(\frac{H}{4 z_0}\right)}  \right] \sech^2 \left(\frac{z}{2 z_0}\right),
\end{equation}
where $\Sigma_{\mathrm{gas}}$ is the initial gas mass surface density, $z_0$ is the scale height of the initial profile, $H=4~\mathrm{kpc}$ is the vertical extent of the box, and the first term in brackets represents the density at the midplane. This density profile acts in conjunction with a $8.6\times10^{-30}~\mathrm{g}~\mathrm{cm}^{-3} $ floor imposed on these initial conditions. We employ a value of $\Sigma_{\mathrm{gas}}=10~\mathrm{M}_{\odot}~\mathrm{pc}^{-2}$ for the initial gas mass surface density, as this surface density falls in the regime where CRs are expected to be dynamically significant \citep{Rathjen2023}, and is near the solar neighborhood value \citep{McKee2015}. Our adopted values for the initial mass of the gas column is similar to the respective solar neighborhood condition simulations of the \textsc{silcc} \citep{Walch2015} and TIGRESS \citep{KimOstriker2017} collaborations. The gas is initialized with solar composition \citep{Asplund2009_solar_comp_crisp}. We use a value of $100~\mathrm{pc}$ for $z_0$, and prescribe an initial uniform temperature of $10^4~\mathrm{K}$ to begin our simulations from conditions corresponding to approximate hydrostatic equilibrium. The addition of radiative processes and magnetic fields violates this ``equilibrium'' initialization. The magnetic field is initialized to have a uniform $P_\mathrm{th}/P_\mathrm{mag}=\beta=1000$, corresponding to an initial peak of $\sim$$0.23~\mathrm{nG}$\footnote{We report magnetic field strengths and equations in the Gaussian system of units.}, and is pointed in the $x$-direction. This seed magnetic field is allowed to amplify as a result of a fluctuating dynamo caused by supernovae-driven turbulence. We neglect the rotational motion of the galactic disk as another source of magnetic field growth via a large-scale dynamo.

In addition to including self-gravity of the gas and gravity between gas and star particles, we include a static external gravitational potential corresponding to an old stellar population of surface density $\Sigma_* = 30~\mathrm{M}_\odot~\mathrm{pc}^{-2}$ \citep[e.g.,][]{Walch2015, KimOstriker2017}. We use a smoothed plane-parallel gravitational acceleration \citep[e.g.,][]{McCourt2012, Ji2018, Butsky2020} of the form
\begin{equation} \label{eq:smoothed_grav_accel}
    g(z) = g_0 \frac{z/a}{[1 + (z/a)^2]^{1/2}},
\end{equation}
where $g_0=2 \pi G \Sigma_*$, and the scale height $a=100~\mathrm{pc}$. We ignore the contribution of the dark matter halo as its vertical component is weak compared to the other gravitational terms. Gravitational forces between the gas and star particles are calculated using the TreePM approach adapted to the geometry of the tallbox \citep[][]{Springel2021}. We use a fixed gravitational softening length of $1~\mathrm{pc}$ for the star particles and adaptive softening for gas particles.

This setup is subject to cooling-induced collapse. One approach to mitigate this unphysical result is to apply turbulence-driving to the gas up until stars have begun forming and providing feedback support \citep[e.g.,][]{KimOstriker2017, Rathjen2023}. We instead include an initial turbulent velocity field of RMS norm $40~\mathrm{km~s}^{-1}$, and continuously inject SNe in the midplane to emulate a feedback-supported steady-state. This ``artificial stirring'' phase takes place over the beginning $50~\mathrm{Myr}$ of the simulations. We sample a hyperbolic tangent function with a scale height of $50~\mathrm{pc}$ about the midplane to stochastically inject SN feedback at a rate corresponding to 1 SN event per every $100~\mathrm{M_\odot}$ of star formation expected from the Kennicutt-Schmidt law \citep{Kennicutt1998_KS_Law} for the initial $10~\mathrm{M}_{\odot}~\mathrm{pc}^{-2}$ gas surface density. This allows the simulations to begin ``natural'' star formation without a catastrophic initial starburst, and is initially equivalent to a ``random-driving'' setup \citep{Simpson2016}. The side-effect of this phase is a strong outflow for the initial $50~\mathrm{Myr}$ and transient outflow for the following $\sim$$30~\mathrm{Myr}$, and is not included in our analysis of the resulting steady-states.

We refine and derefine gas cells following the target-mass prescription in \citet{Springel2010_AREPO} with resolution $M_{\mathrm{target}}=10~\mathrm{M}_\odot$ -- a factor of 100  smaller than in \tttf. In addition, we impose a volume ceiling on cell sizes. This acts to retain cold gas and prevent hot gas cells from becoming excessively large. Super-Lagrangian refinement of this type has become more popular in global simulations \citep{Hummels2019_super_Lagrangian_refinement, Suresh2019_Super_Lagrangian_refinement, vandeVoort2019_refinement, ThomasPfrommerPakmor2023_Wave_Dark, TimonGlobal2024, Rey2024_refinement}. This refinement strategy allows the \textsc{Arepo} moving-mesh structure to retain advantages of a finite mass approach in the ISM and a spatially-resolved, finite volume approach in the wind. We approach this by first applying a maximum cell volume $V_{\mathrm{max}}=10^{-6}~\mathrm{kpc}^3$ and minimum cell volume $V_{\mathrm{min}}=10^{-11}~\mathrm{kpc}^3$. Assuming Voronoi cells are approximately spherical with radius $\Delta x$, this gives a preliminary $\Delta x_{\mathrm{max}}\approx 6.2~\mathrm{pc}$ and $\Delta x_{\mathrm{min}}\approx0.13~\mathrm{pc}$ (although this minimum radius is only realized in star-forming cells). To prevent derefinement of cold gas in the chaotic galactic wind, we also impose a stronger target radius criterion. In practice, this acts as a maximum radius for low mass cells. We define the target radius to vary with the cell density as
\begin{equation} \label{eq:special_volume_refinement}
    \frac{\Delta x_{\mathrm{target}}}{\mathrm{pc}} = \max{\left\{-\log{\left(\frac{n_{\mathrm{H}}}{\mathrm{cm}^{-3}}\right)+1},3\right\}},
\end{equation}
giving an effective resolution of $\Delta x = 3~\mathrm{pc}$ for cold gas in the wind.

We carry out four simulations, each with a progressive modeling addition. The first simulation contains no CRs (``\mhdX''). The second run includes CRs injected at SNe, but considers CR advection as the only CR transport mechanism (``\advX''). Instead of utilizing the two-moment solver from \citet{ThomasPfrommer2022_CRHD_AREPO_2}, the \advX case applies the one-moment advective flux solver from \citet{Pfrommer2017_AREPO_CR}. The third simulation includes CRs with the \crmhd model and only NLLD of Alfvén waves (``\nlX''). The fourth and final ``full-physics'' simulation uses the \crmhd model with NLLD and IND (``\inX''). \nlX and \inX, as a pair, are referred to as the ``\crmhd'' cases, for their inclusion of the two-moment \crmhd model \citep{ThomasPfrommer2019_CRHD}. Aside from the implementation or absence of CRs, these four simulations have identical setups. The simulations are run for a total of 250 Myr to ensure that the they attain a meaningfully long steady-state outflow for analysis. We omit the bulk of the analysis of the advection-only case due to star formation results discussed in Section~\ref{sec:star_formation}. The other three simulations will be interchangeably referred to as the ``primary'' or ``star-forming'' runs. Due to the fairly idealized nature of these runs, we also caution against making absolute comparisons to observed outflows, and instead focus on the relative differences between these simulations.

\section{Structure and Profiles}
\label{sec:structure_and_profiles}

\begin{figure*}
\includegraphics{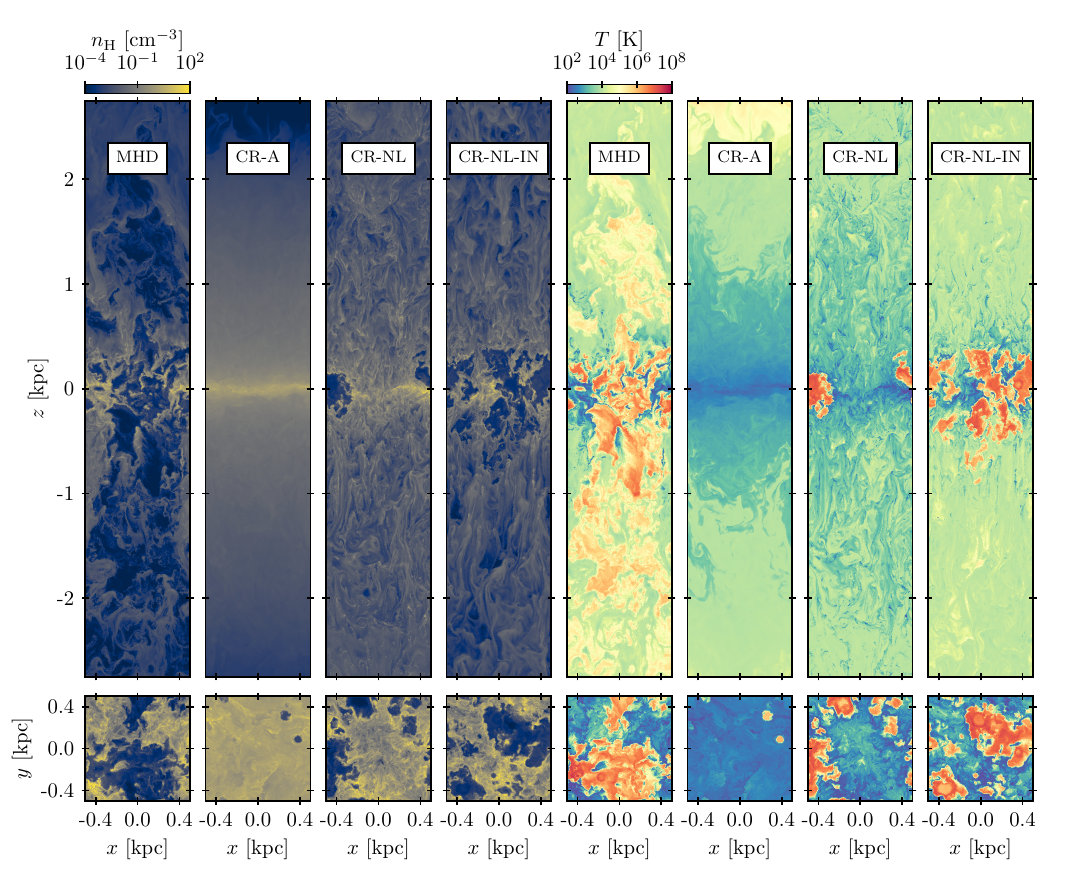}
\caption{Vertical (top) and horizontal (bottom) slices through the domains. Total hydrogen number density is shown on the left, and gas temperature on the right. From left to right, we show the \mhdX case, \advX case, \nlX case, and the \inX case. All snapshots are taken at $t=250~\mathrm{Myr}$, representative of a steady-state wind. The resulting gas is hotter in the pure MHD case and cooler in the cases with CRs.
\label{fig:dens_temperature_slices}}
\end{figure*}

\begin{figure}
\includegraphics[width=\linewidth]{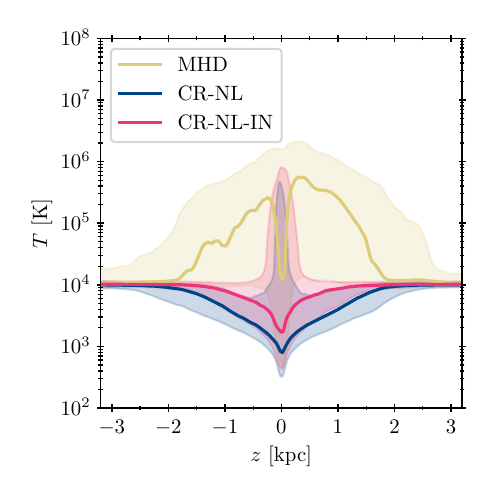}
\caption{Vertical temperature profiles through the domains of the three primary simulations, averaged over the steady-state wind for $112~\mathrm{Myr}<t<250~\mathrm{Myr}$. The volume-weighted median is represented by the thick line, and the shaded regions enclose the $25^{\mathrm{th}}$ and $75^{\mathrm{th}}$ percentiles. The wind is preferentially hotter than $10^4~\mathrm{K}$ in the case without CRs and colder in the \crmhd cases.
\label{fig:temperature_profiles}}
\end{figure}

In Figure~\ref{fig:dens_temperature_slices} we show vertical and horizontal slices through each domain. We display the total hydrogen number density and the temperature at $t=250~\mathrm{Myr}$, representative of the steady-state for each case. We produce a full variety of gas phases under each prescription, from very cold molecular gas to hot diffuse gas. 

Focusing first on the midplane slices (Figure~\ref{fig:dens_temperature_slices} lower), we see morphological signatures of a feedback-regulated, dynamic ISM in the three primary simulations. The cool/warm gas fills most of the area and is interspersed with cold, dense pockets and hot, sparse bubbles. The \mhdX case has the most hot gas, followed by \inX and \nlX. This is likely a consequence of different star formation rates and thus different SN rates, as the SNe are expected to strongly alter the hot content of the ISM \citep{Simpson2023}. Aside from the visual differences in the hot gas volume filling fraction, the midplanes of the three primary cases do not show any striking differences. The case with purely advective CRs shows a primarily cold midplane with two hot SNRs, suggesting a very low SFR.

In contrast to the midplane slices, the wind slices show very clear differences. The \mhdX scenario produces a primarily warm/hot ($\sim$$10^{4}~\mathrm{K}$ to $\sim$$10^{6}~\mathrm{K}$) wind with a handful of dense cloudlets. The majority of the wind is low-density (expected of warm/hot gas). The results from the \advX simulation reveal a very smooth midplane filled with cold, dense gas and no evidence of feedback activity.

The \nlX and \inX cases show visually unique winds in comparison with each other and the \mhdX case. The \nlX simulation produces a wind filled with cold, dense cloudlets surrounded by warm ($\sim$$ 10^{4}~\mathrm{K}$) gas. There is very little hot gas in this wind. In contrast, the full-physics case produces a primarily warm wind by volume, interspersed with pockets of hot gas and cold cloudlets. This case also shows hot gas ejecting from the midplane, likely due to a SN superbubble breakout. The differences between these two \crmhd cases could be attributed to both differences in amount of feedback (i.e., different SFRs) and IND resulting in CRs decoupling from cold gas in the \inX case.

The vertical volume-weighted temperature profiles for the star-forming cases are shown in Figure~\ref{fig:temperature_profiles} (The omission of the advective case is discussed in Section~\ref{sec:star_formation}.). We show the median values as well as the $25^{\mathrm{th}}$ and $75^{\mathrm{th}}$ percentiles of fluctuations as a shaded region, with data from $112~\mathrm{Myr}$ to $250~\mathrm{Myr}$. We find that the cases with CRs tend to result in cooler gas everywhere compared to the case without CRs. This matches what is seen in the slices from Figure~\ref{fig:dens_temperature_slices}, which show temperature differences spatially. The pure MHD simulation has a wide distribution of gas temperatures in the midplane surrounded by extra-planar outflowing SN-heated gas. The cases with CRs also show a cold midplane, but their midplanes are instead surrounded by cold gas in the \nlX case or warm gas in the \inX case. The difference in temperatures produced in the respective winds is interpreted as a signature of the feedback experienced by each galaxy, both in the magnitude of the SFR and in the CR transport model. We characterize the pure MHD case as producing a primarily hot wind ($\sim$$10^{6}~\mathrm{K}$), the \nlX case producing a cool wind ($< 10^{4}~\mathrm{K}$), and the \inX case producing a warm wind ($\sim$$10^{4}~\mathrm{K}$).

The case without CRs also shows a minor asymmetry, with a slightly hotter wind in the $+z$-direction as compared to the $-z$-direction. This can be attributed to the aggressive, sporadic hot wind that is central to the thermally-driven wind paradigm. Strong outbursts associated with superbubble breakouts generally occur through a channel in which material has been cleared, and the over-pressurized material may evacuate without much resistance. This may stochastically show favor to a direction in which material has already been cleared, but would likely not produce a significant asymmetry in a longer, time-averaged analysis. Instantaneous asymmetries can be observed in a starburst scenario, where superbubble breakouts are more common \citep[e.g.,][]{ShopbellJBH1998_asymmetric_wind_m82}.

\begin{figure*}
\includegraphics{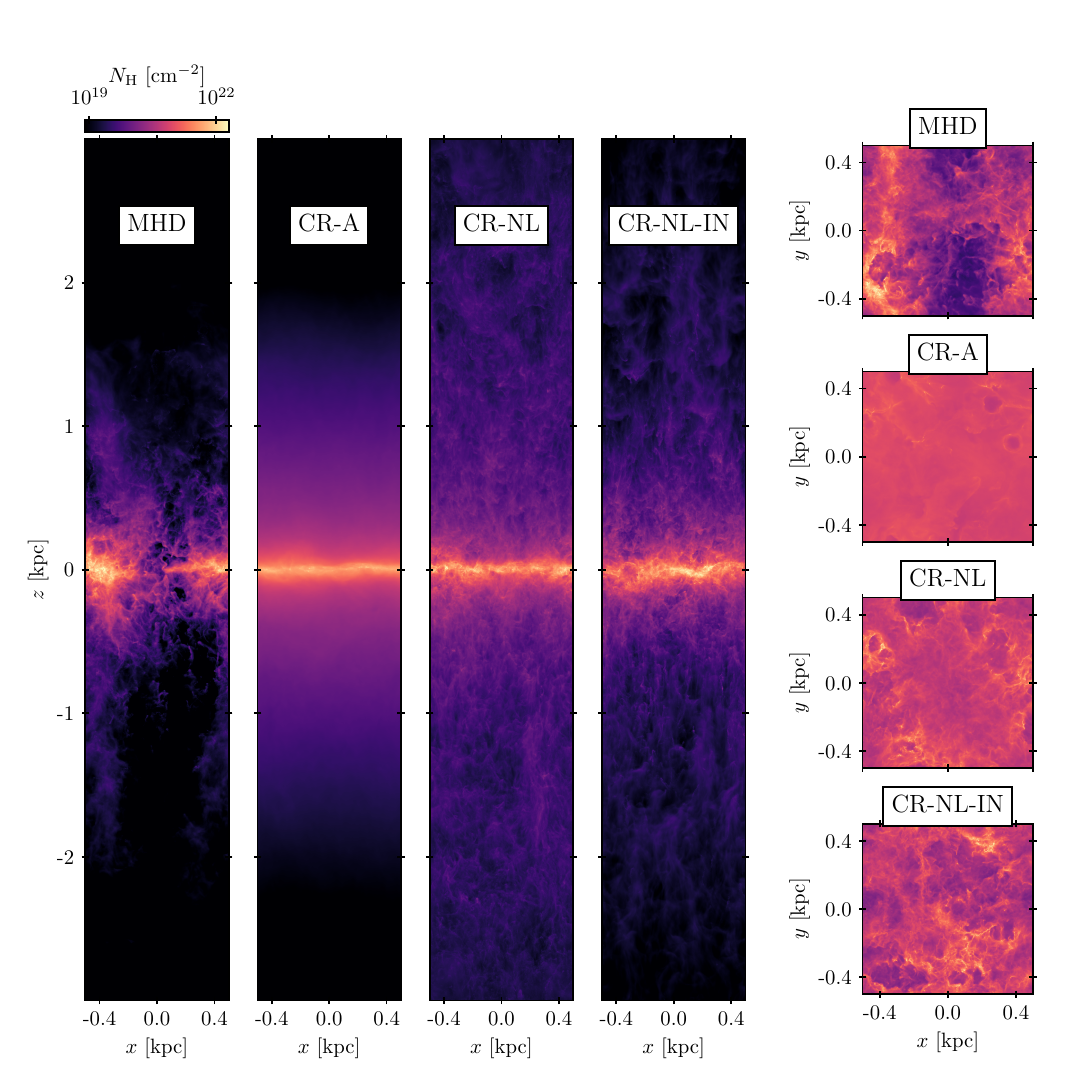}
\caption{Edge-on (leftmost) and face-on (rightmost) projections of the total hydrogen number density through the simulation domains. All snapshots are taken at $t=250~\mathrm{Myr}$ as in Figure~\ref{fig:dens_temperature_slices}. From left to right in the edge-on plots and from top to bottom in the face-on plots, we show the \mhdX case, \advX case, \nlX case, and \inX case. The \crmhd cases produce a visually denser wind and ISM compared to the pure MHD case.
\label{fig:dens_projections}}
\end{figure*}

In Figure~\ref{fig:dens_projections}, we show edge-on and face-on column densities of the total amount of hydrogen in each simulation at $t=250~\mathrm{Myr}$, representative of a steady-state. The different physical prescriptions produce immediately visible differences in both the edge-on and face-on morphology of each simulation. Focusing first on the edge-on projections, we see that the winds vary in both clumpiness and overall density. The simulation without CRs displays an excavated wind around $x=0~\mathrm{kpc}$ with extra-planar gas in ``columns'' near $x=\pm0.5~\mathrm{kpc}$. This reflects the structural inhomogeneity imposed by thermally-driven winds, where superbubble breakouts produce volume-filling, hot, tenuous gas. The \nlX and \inX cases are distinct from the purely MHD case in that they contain more extra-planar material in the form of dense cloudlets and filaments. The dense extraplanar material in the \nlX case is nearly volume-filling. This is in contrast to the \inX case, where the clumps can appear separated by underdense regions. The vertical extent of the gas in each projection reflects the relative differences between the temperature profiles in Figure~\ref{fig:temperature_profiles} and the edge-on slices in Figure~\ref{fig:dens_temperature_slices}. The case without CRs has a hot, low density wind surrounding the midplane, which appears empty in projection. The \nlX case has a cold, dense wind with high column densities and therefore appears vertically extended in projection. The full-physics case is an intermediate between the \nlX and \mhdX cases: The cold gas dominates the column density in projection and therefore dictates its vertical extent, but is still interlaced with warm/hot gas that appears as visually excavated regions.

The midplanes of both \crmhd cases remain intact, in contrast to the broken midplane of the pure MHD case. The advective case is visually distinct in that it shows very little morphological evidence of SN feedback. It instead shows a smooth, vertically-extended gas profile. These distinctions are likely attributed to differences in stellar feedback activity, explored in Section~\ref{sec:star_formation}.

Looking now at the face-on projections in Figure~\ref{fig:dens_projections}, we see the midplane differences resulting from the different feedback prescriptions. Face-on morphology is expected to be strongly affected by recent stellar activity, and therefore can vary over time. We present these snapshots as representative of a fluctuation within a steady-state. The \mhdX case shows significant spatial variance resulting from SNe, reminiscent of the face-on profiles produced in the solar neighborhood TIGRESS simulations \citep{KimOstriker2017}. The \crmhd cases both show evidence of SN feedback, but do not produce voids that are as low in column density as in the \mhdX case. The \inX case notably shows a greater variety in column densities as compared to the \nlX case, possibly owing to an increased SN rate, placing \inX as an intermediate between the \mhdX and \nlX cases. The simulation with purely advective CRs again shows very little morphological evidence of SNe. The difference between the MHD and the \crmhd cases seen in this work is not found in \tttf; however \tttf analyze the galaxies at a steady state around $t=1~\mathrm{Gyr}$, where the SFRs have converged, which is much later than in this analysis. The reduction in SFR by CRs observed in our simulations, but not in those of \tttf, is likely attributable to a combination of our earlier analysis time and superior mass resolution. We note also that the face-on morphology of disk galaxies is sensitive to structures such as spiral arms \citep{Kormendy2004_disk_spiral}, which are not captured in this work.

\begin{figure}
\includegraphics{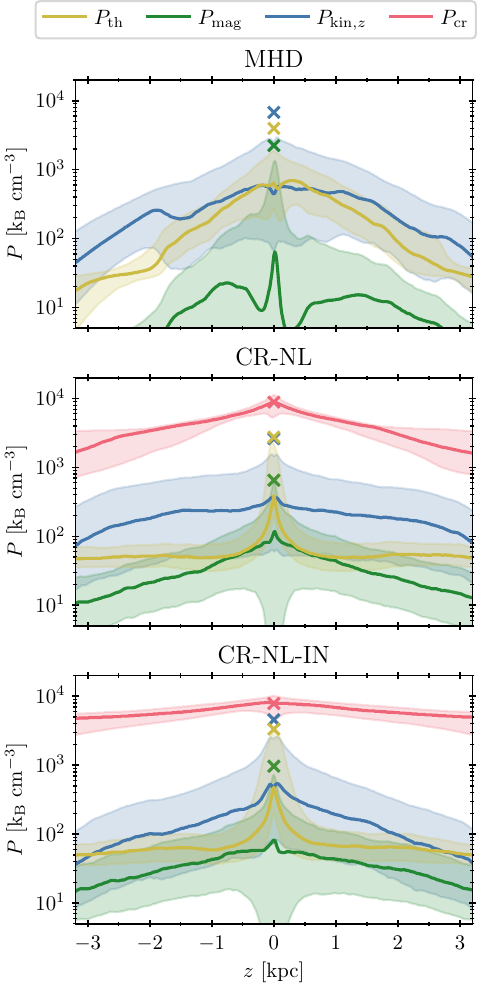}
\caption{Vertical pressure profiles through the domains of the \mhdX simulation (top panel), \nlX simulation (middle panel), and \inX simulation (bottom panel). We include the thermal pressure $P_\mathrm{th}$, the total magnetic pressure $P_{\mathrm{mag}}=(B_x^2+B_y^2+B_z^2)/8\pi$, the vertical ram pressure $P_\mathrm{kin,z}=\rho v_z^2$, and the CR pressure $P_\mathrm{cr}$, with color-coding displayed at the top of the figure. The median value is displayed as a thick line, and the $25^{\mathrm{th}}$ and $75^{\mathrm{th}}$ volume-weighted percentiles bound the shaded area. The volume-weighted averages (as opposed to the medians) at the midplane are marked with an ``$\times$''. Thermal and kinetic pressures are in approximate equipartition in the midplane in all cases. Magnetic pressure is lower but comparable. CR pressure dominates at all heights in both \crmhd cases, with a slightly flatter profile in the full-physics run. 
\label{fig:pressure_profiles}}
\end{figure}

In Figure~\ref{fig:pressure_profiles}, we show the vertical pressure profiles for the three primary cases. We show the median pressure values along with shading corresponding to $25^{\mathrm{th}}$ and $75^{\mathrm{th}}$ percentiles as weighted by volume, over $112~\mathrm{Myr}<t<250~\mathrm{Myr}$. We plot the thermal pressure $P_\mathrm{th}$ alongside the total magnetic pressure $P_{\mathrm{mag}}=(B_x^2+B_y^2+B_z^2)/8\pi$, the ram pressure in the $z$-direction $P_\mathrm{kin,z}=\rho v_z^2$, and the CR pressure $P_\mathrm{cr}$.

We find near equipartition between thermal, vertical ram, and total magnetic pressures in the midplane, with the magnetic pressure slightly below the other two. For the cases with CRs, we find that the CR pressure dominates at all heights. This is consistent with post-processed tallbox work \citep{Armillotta2021, Armillotta2024} and mostly consistent with \tttf~when neglecting vertical magnetic stress.

We choose to display the magnetic pressure as opposed to the vertical magnetic stress. \tttf~and tallbox simulations with shearing terms \citep{Vijayan2020, Armillotta2022} show that the vertical magnetic stress can reach equipartition near the midplane and become negative outside. Our simulations instead produce a vertical magnetic stress that is very small near the midplane and preferentially negative elsewhere. This implies that $B_z^2 \gtrsim B_x^2 + B_y^2$, or that the magnetic field orientation is primarily vertical as opposed to planar--a natural result of our ``galactic wind tunnel'' setup in which the primary mode of magnetic field amplification is through flux-frozen outflows. Without imposed shearing or natural differential motion to produce toroidal structure \citep[e.g.,][]{ThomasPfrommerPakmor2023_Wave_Dark}, we underdevelop the planar magnetic field. The behaviors of the magnetic stress away from the midplane (where wind-driving occurs) are broadly consistent with the other work, as the vertical motions are dominant and should certainly dictate the magnetic topology. Therefore, this effect should not heavily impact the outflow or the relative differences between the simulations. We reiterate that our focus is on the relative comparisons between various CR transport physics cases rather than on including all relevant physics and obtaining results that are correct in absolute terms.

Although we still prioritize relative comparisons between the simulation cases, we compare with the observed solar-neighborhood CR pressure value to provide links to similar work \citep[e.g.,][]{Armillotta2021}. For reference, we take the solar-neighborhood value of CR energy density to be $1.9~\mathrm{eV}~\mathrm{cm}^{-3}$ \citep{Grenier2015_CR_Nine_Lives}, corresponding to an isotropic pressure of approximately $7350~\mathrm{k_B}~\mathrm{cm}^{-3}$. Both \crmhd cases are consistent with this midplane value within the $25^{\mathrm{th}}$ and $75^{\mathrm{th}}$ percentiles, although our median value in each case is slightly above this solar-neighborhood value. Our midplane value of CR pressure in our \inX case is higher than $\sim$$1.8\times10^3 \; k_\mathrm{B} \; \mathrm{cm}^{-3}$ reported by \tttf. This difference is likely due to the two times higher SFR in our simulations compared to that reported by \tttf. Since in \tttf~simulations, the SFR decreases by a factor of $\sim$$2$ between $t\sim 200~\mathrm{Myr}$ and  $t=1~\mathrm{Gyr}$, the difference in the CR energy density that we and \tttf~measure is likely due to the fact that we analyze data corresponding to earlier evolutionary time ($t\sim200~\mathrm{Myr}$) than \tttf.

The CR energy density calculated by \citet{Grenier2015_CR_Nine_Lives} is about $0.5~\mathrm{eV}~\mathrm{cm}^{-3}$ greater than the value reported in \citet{Draine2011} and \citet{RydenPogge2021}, and $0.9~\mathrm{eV}~\mathrm{cm}^{-3}$ greater than the value inferred from Voyager 1 data in \citet{Cummings2016_voyager_CRs}. Our solar-neighborhood CR energy density is still consistent with the value in \citet{Draine2011}, but our midplane CR energy density is always above the value reported in \citet{Cummings2016_voyager_CRs}. Due to the discrepancies between the model-dependent values of CR energy density in the solar neighborhood reported by \citet{Draine2011}, \citet{Grenier2015_CR_Nine_Lives}, and \citet{Cummings2016_voyager_CRs}, we place little weight on matching our CR energy density values with these specific observational results. Additionally, the correct value of the solar neighborhood energy density may not in principle be representative of typical conditions along the solar circle.

The calculated value of the midplane pressures depends strongly on how the value is determined, either by median or by averaging. We find that the volume-weighted average can be more than an order of magnitude higher than the median. This is purely a statistical effect. The pressures at any given height generally span an order of magnitude. Because the $25^{\mathrm{th}}$ and $75^{\mathrm{th}}$ percentiles of these pressure distributions are separated from the median by similar dex, we can qualitatively discuss the pressures as if they constitute a log-normal distribution. For a log-normal distribution, the mean is always higher than the median. This can be understood by re-mapping the distribution to linear space, where a log-normal distribution will have a significant ``tail'' of high values (depending on the width of the log-normal distribution), and the mean will be strongly determined by these high values. Because of this statistical property, the mean midplane thermal, magnetic, and kinetic pressures are significantly separated from their respective medians. The midplane CR pressure is mostly insensitive to this statistical effect because of the small variance of CR pressures.

When looking at the midplane average pressures for the full-physics \inX case in Figure~\ref{fig:pressure_profiles}, we see that the CR pressure is the dominant midplane pressure component. Our midplane magnetic pressure is lower than the observed solar neighborhood value, likely owing to our omission of shearing terms, which would amplify the magnetic field via an additional large-scale dynamo. 

Analyzing the midplane thermal pressure requires a careful approach. The volume-weighted average midplane thermal pressure in the \inX case corresponds to an energy density of approximately $0.43\;\mathrm{eV}\;\mathrm{cm}^{-3}$, which is apparently consistent with the solar neighborhood value of $0.4\;\mathrm{eV}\;\mathrm{cm}^{-3}$ \citep{Draine2011,RydenPogge2021}. However, the value of $0.4\;\mathrm{eV}\;\mathrm{cm}^{-3}$ is based on analysis of intervening neutral carbon absorption features in observed stellar spectra \citep{Jenkins2011_ISM_p_therm}. The analysis in \citet{Jenkins2011_ISM_p_therm} focuses on cold ($T\sim80\;\mathrm{K}$) gas, with a preference for higher densities. Using field stars (effectively a volume-weighted sample) with a preference towards higher densities produces a net mass-weighted sampling. Therefore, as a more apt comparison, we calculate the mass-weighted average thermal pressure for cold ($T<184\;\mathrm{K}$) midplane ($|z|<200\;\mathrm{pc}$) gas. For the values of $\log \left(\langle P_\mathrm{th} \rangle \left[k_\mathrm{B}\;\mathrm{cm}^{-3} \right] \right)$, we find $3.16$, $3.07$, and $3.10$ for the \mhdX, \nlX, and \inX cases respectively. These are all slightly lower than the value of $3.58$ from \citet{Jenkins2011_ISM_p_therm}. Additionally, if we instead calculate the neutral-carbon-mass-weighted mean with a lower bound of $0.2$ on the molecular hydrogen fraction $x_{\mathrm{H}_2}$, we find values of $3.26$, $3.13$, and $3.18$ for the three cases.

It is unclear what is the most correct comparison with observations. \citet{Kim2023_tigress_NCR}, for example, compare their mass-weighted pressure distributions with the value from \citet{Jenkins2011_ISM_p_therm}, and find good agreement. However, preceeding simulations with a simpler ISM model from \citet{KimOstriker2015} underestimate the ISM thermal pressure. It is plausible that our underestimation of the midplane thermal pressure is due to missing physics (such as explicit radiative transfer and photo-ionization), in addition to uncertainties from comparing with observations.

\section{Quantifying Feedback}
\label{sec:star_formation_and_outflows}

In this section, we explore the effects of feedback on star-formation and outflows. We quantify the differences between the simulations by comparing star formation rate, cumulative stellar mass formed, mass loading, and energy loading.

\subsection{Star Formation}
\label{sec:star_formation}

\begin{figure}
    \includegraphics[width=\linewidth]{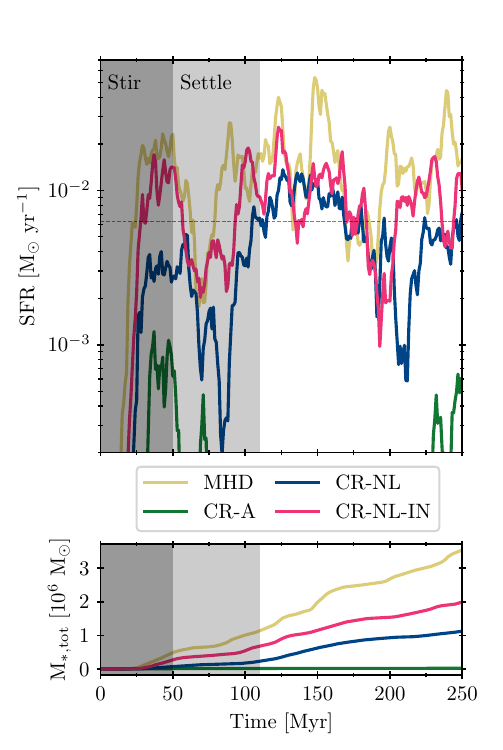}
    \caption{Star formation rate (SFR; top) and cumulative star formation (bottom) for each simulation. We shade the artificial stirring period that takes place over the initial $50~\mathrm{Myr}$ in dark gray. We also shade the following transient outflow period of $\sim$$ 60~\mathrm{Myr}$ in light gray. The dashed line in the SFR corresponds to the Kennicutt-Schmidt rate for the initial gas surface density \citep{Kennicutt1998_KS_Law}. While CRs tend to decrease the SFR, IND reduces the effect of CRs.
    \label{fig:sfr_sftot}}
\end{figure}

Figure~\ref{fig:sfr_sftot} shows the instantaneous global star formation rate as a function of time, as well as the cumulative mass of stars formed. We find that, after an appropriate time has passed to disinfect from initial transients, the simulations each reach an approximately steady SFR. We note the aggressive quenching of the purely advective CRs case. In this case, there is no CR transport relative to the gas. CRs end up stuck in the midplane, preventing star formation and inflow of new material toward the galactic disk midplane. During this long stretch of time without star formation, CR energy is primarily lost through hadronic and Coulomb interactions. Eventually enough CR pressure is dissipated to allow for another star formation event near the end of the simulated time period. Due to this lapse in star formation, mass and energy loading factors are undefined and there is no meaningful outflow to be studied. We choose to neglect the advective CRs case for the remainder of this work. The other cases provide sufficient room for insight.

We shade out the initial stirring phase (discussed in Section~\ref{sec:crisp_ism_model}), as well as an additional $\sim$$60~\mathrm{Myr}$ afterwards. The initial starburst can produce an aggressive outflow that is more energetic than the steady-state outflow seen in the rest of the simulation \citep[cf.\ also][]{Rathjen2023}. We choose not to analyze the results until these transient effects have subsided to ensure the simulation is more representative of a steady-state galaxy.

We clearly see that the case without CRs produces the most stars and tends to have the highest instantaneous star formation rate. The \nlX case forms less mass in stars than \inX. This difference in SFR contributes to the morphological differences discussed in Section~\ref{sec:structure_and_profiles}. More star formation leads to more frequent SN feedback and a hotter ISM--a conclusion similar to the one obtained in the context of studies of peak-driven SNe in the ISM \citep{Simpson2023}. The ordering of the star formation rates also coincides with the ordering of the wind temperatures, demonstrating that small differences in feedback can produce significant differences in the content of the galactic wind.

\begin{figure}
    \includegraphics[width=\linewidth]{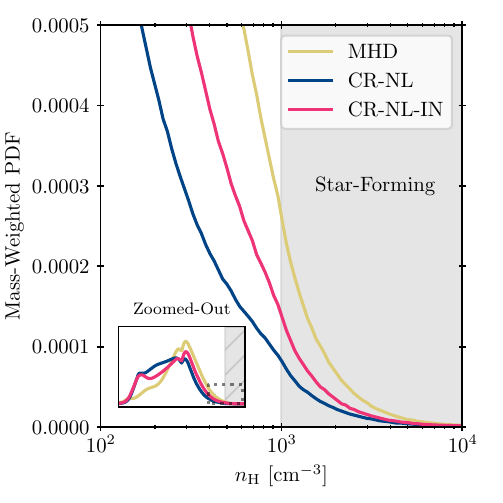}
    \caption{Mass-weighted density PDF near star-forming densities for the three star-forming cases, with an inset (lower left) to show the same PDF for all densities, averaged for $112~\mathrm{Myr}<t<250~\mathrm{Myr}$. The shaded region shows the densities for which gas is designated to be star-forming. The trend in the amount of star-forming gas is mirrored by the SFR in Figure~\ref{fig:sfr_sftot}.
    \label{fig:sf_mass_phase}}
\end{figure}

In Figure~\ref{fig:sf_mass_phase}, we show the mass-weighted distribution of gas density throughout the entire domain, averaged over the steady-state analysis period. The runs with more gas at star-forming densities
match the runs with more star formation in Figure~\ref{fig:sfr_sftot}. That is, the \mhdX case preferentially contains more star-forming gas, the \nlX case contains the least star-forming gas, and the \inX case lies in-between the other two. This is a necessary requirement to produce the differences in SFR. The underlying physical assumption for our star formation model is that stars are formed above a threshold density. An increase in the proportion of mass above this density threshold directly means an increase in the SFR. The differences between the three analyzed cases can be easily explained by the presence of CRs and their transport physics. Without CRs, there is less pressure support in the ISM. Consequently, more gas is able to collapse to star-forming densities. When CRs are added, they prevent gas from cooling and reaching star-forming densities. IND causes CRs to decouple from this collapsing gas, permitting the formation of stars. This argument relies on the assumption that there is no additional CR confinement by extra-planar hot gas \citep{Armillotta2021}. However, \citet{Armillotta2021} draw the conclusion that CRs are confined by extra-planar gas based on post-processed simulations of CR transport in a star-forming galactic disk without self-consistent CR feedback. Figures~\ref{fig:dens_temperature_slices} and \ref{fig:temperature_profiles} show that the extra-planar region contains less hot gas with the inclusion of CR feedback. Therefore, it is likely that CRs in our self-consistent \inX simulation are less confined to the midplane than CRs in \citet{Armillotta2021}.

\subsection{Outflows and loading factors}
\label{sec:loading_factors}

\begin{figure}
    \includegraphics[width=\linewidth]{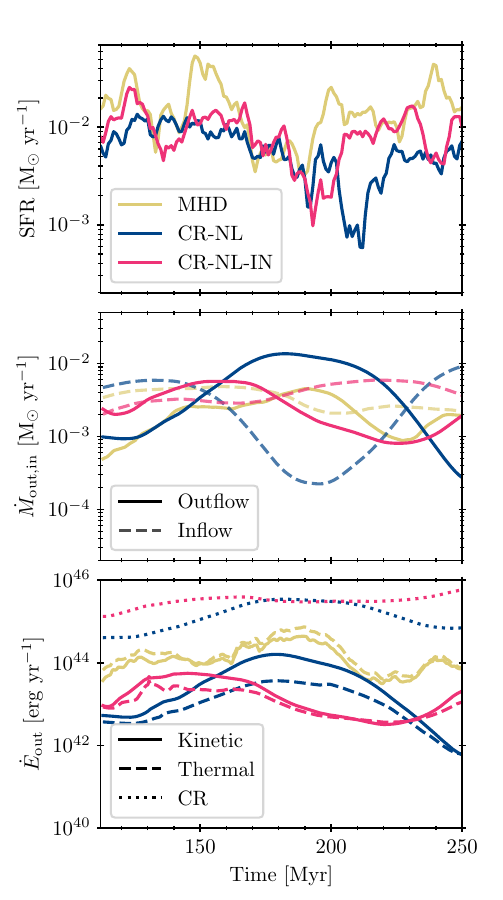}
    \caption{Instantaneous SFR (upper panel), outward (solid) and inward (dashed) mass flux (middle panel), and outward energy flux (lower panel) for the steady-state outflow period and $1~\mathrm{kpc}<|z|<3~\mathrm{kpc}$ to represent the spatial body of the wind. The mass outflow and star formation rates are all fairly comparable between the cases. In the \crmhd cases, energy is primarily carried by CRs.
    \label{fig:sfr_mdot_edot_time}}
\end{figure}

In Figure~\ref{fig:sfr_mdot_edot_time} we show the star formation rates alongside the mass outflow/inflow and energy outflow rates averaged within $1~\mathrm{kpc}<|z|<3~\mathrm{kpc}$. We define
\begin{equation} \label{eq:mass_flux}
    \dot{M}_{\mathrm{out,in}} = \frac{1}{\Delta z} \sum_i M_i v_{\mathrm{out,in,}\,i},
\end{equation}
to be the outward/inward mass flux within a region of height $\Delta z$ and $M_i$ is the mass in computational cell $i$. The outward/inward velocity $v_{i,\mathrm{out,in}}$ is defined as $v_z \sign(z)$, and each cell is separated into the inward/outward categories based on the sign of this quantity (meaning the summation in Equation~\eqref{eq:mass_flux} will always be a sum of like terms). We similarly define the energy outflow rate as
\begin{equation} \label{eq:energy_flux}
    \dot{E}_{\mathrm{out}} = \frac{1}{\Delta z}\sum_i E_i \velsym_{\mathrm{out,}\,i},
\end{equation}
where $E_i$ is the energy quantity of interest. We take
\begin{align}
    E_{\mathrm{kin,}\,i} &= \frac{M_i}{2} \velsym_i^2 \quad \text{as the kinetic and} \\
    E_{\mathrm{th,}\,i} &= V_i (\varepsilon_{\mathrm{th},i} + P_{\mathrm{th,}\,i})  \quad \text{as the thermal} 
\end{align}
energies, where $\varepsilon_{\mathrm{th,}\,i}$ is the thermal energy density, $P_{\mathrm{th,}\,i}$ is the thermal pressure, and $V_i$ is the volume of the Voronoi cell. For the \crmhd cases, the CR energy outflow is also augmented by the flux along magnetic field lines. This results in a modified CR energy outflow rate
\begin{equation} \label{eq:energy_flux_cr}
    \dot{E}_{\mathrm{out,cr}} = \frac{1}{\Delta z}\sum_i E_{\mathrm{cr,}\,i} \velsym_{\mathrm{out,}\,i} + F_{\mathrm{cr,}\,i},
\end{equation}
where the advective CR energy and momentum flux terms are
\begin{align} \label{eq:cr_flux_in_energy}
    E_{\mathrm{cr,}\,i} &= V_i (\varepsilon_{\mathrm{cr,}\,i} + P_{\mathrm{cr,}\,i}),\\
    F_{\mathrm{cr,}\,i}&=V_i f_{\mathrm{cr,}\,i}
    (\bm{b}
    \bm{\cdot} \bm{e}_z) \; \sign(z),
\end{align}
where the CR energy flux density is $f_{\mathrm{cr,}\,i}$ and $(\bm{b}\bm{\cdot}\bm{e}_z) \sign(z)$ is the component of the magnetic field directed away from the midplane.

We see that all of the mass outflow and inflow rates generally lay between $10^{-3}~\mathrm{M_\odot}~\mathrm{yr}^{-1}$ and $10^{-2}~\mathrm{M_\odot}~\mathrm{yr}^{-1}$, with wide variations over timescales of $50~\mathrm{Myr}$. The outflow and inflow rates are typically anti-correlated. However, this anti-correlation is present in the statistical sense as, at any given time, both outflows and inflows are present forming a ``fountain flow.'' We discuss this process further in Section~\ref{sec:outflow_content}.

The \nlX case shows a notable excess of mass outflow from $t\sim160~\mathrm{Myr}$ to $t\sim220~\mathrm{Myr}$, during which the star formation rate is declining. In this time period, we see that the mass outflow rate slightly exceeds the star formation rate, meaning that more gas is instantaneously being processed by the outflow than through star formation. This is the only physics case to show this sustained excess.

The energy outflow rates show more significant differences between the physics cases. The \mhdX case has the highest kinetic and thermal energy outflow rates, with a slight preference towards thermal energy. This is a result of the hot, SN-driven wind produced in this case. The outflowing gas is low-density, translating to a low mass contribution to the kinetic energy term. The high temperature contributes to the thermal energy term, producing the disparity between the two $\dot{E}_{\mathrm{out}}$ values. In contrast, the \nlX case develops a preferentially kinetic wind. This is explained as the opposite of the \mhdX case: the wind is cold and dense, increasing the mass term in the kinetic outflow and decreasing the temperature term in the thermal outflow. The full-physics \inX outflow has a slight preference towards kinetic energy, but less so than the \nlX case, reflecting a primarily warm wind.

Figure~\ref{fig:sfr_mdot_edot_time} shows that the values of CR energy output are always above the thermal and kinetic values. This is consistent with the results presented by \tttf, and is an important piece of the CR-driven wind paradigm, discussed further in Section~\ref{sec:discuss_t24}. The steadiness of the CR energy fluxes is likely related to the steadiness of the CR pressure profiles in Figure~\ref{fig:pressure_profiles}.

\begin{figure}
    \includegraphics[width=\linewidth]{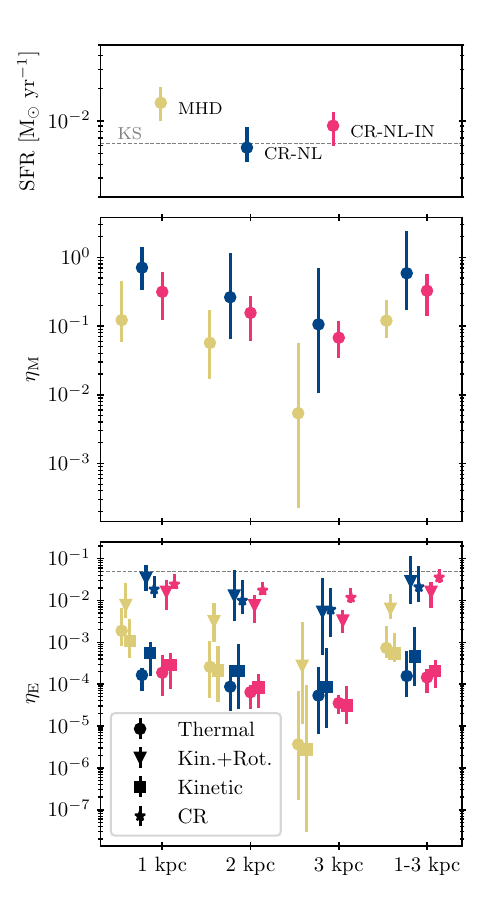}
    \caption{Time-averaged values for the SFR (top), mass loading factor (middle), and energy loading factors (bottom), with points displayed at the median value and error bars representing $25^{\mathrm{th}}$ and $75^{\mathrm{th}}$ percentiles of fluctuations over $112~\mathrm{Myr}<t<250~\mathrm{Myr}$. For the loading factors, we include samplings of slices at $|z|=1$, $2$, and $3~\mathrm{kpc}$, as well as a volume sample of $1~\mathrm{kpc}<|z|<3~\mathrm{kpc}$. The horizontal dashed line at $5\%$ in the energy loading plot represents the injection efficiency of CRs. The \crmhd simulations sustain an outflow to a greater height than the pure MHD simulation. Including the rotational component in kinetic energy significantly boosts the energy loading, especially for the \crmhd cases.
    \label{fig:height_avg_etas_and_sfr}}
\end{figure}

Figure~\ref{fig:height_avg_etas_and_sfr} shows the time-averaged SFR, mass loading, and energy loading factors as a function of height. The mass loading factor is defined as
\begin{equation} \label{eq:mass_loading_eta}
    \eta_{\mathrm{M}} = \dot{M}_{\mathrm{out}}/\mathrm{SFR},
\end{equation}
where $\dot{M}_{\mathrm{out}}$ is defined in Equation~\eqref{eq:mass_flux}. The mass loading factor can either be thought to represent the efficiency of feedback in driving outflows or the ratio of the importance of outflows to star-formation in processing gas. A high value of $\eta_{\mathrm{M}}$ means that stellar activity is very efficient in driving outflows, and that outflows are processing baryons more efficiently than star formation does. We define the energy loading in a similar manner as
\begin{equation} \label{eq:energy_loading_eta}
    \eta_{\mathrm{E}} = \left(\frac{10^{51}~\mathrm{erg}}{100~\mathrm{M}_{\odot}} \; \mathrm{SFR}\right)^{-1} \dot{E}_{\mathrm{out}},
\end{equation}
where the conversion factor in front of the $\mathrm{SFR}$ is an approximate global SN energy injection rate (ignoring the physical delay of feedback), and $\dot{E}_{\mathrm{out}}$ is defined in Equation~\eqref{eq:energy_flux}. The energy loading factor is therefore the efficiency of stellar feedback that contributes energy to the outflow.

In addition to the energy loading factors, we consider another quantity, which we designate ``Kin.+Rot.'' The value of kinetic $\dot{E}_{\mathrm{out}}$ in both global simulations and observations is essentially defined as $\frac{1}{2} \rho \velsym_{\mathrm{tot}}^2 \velsym_{\mathrm{out}}$, where $\velsym_{\mathrm{out}}$ is the velocity in the outflow direction as expected, and $\velsym_{\mathrm{tot}}$ is the total velocity of the gas. If the reference velocity is taken to be the rest-frame of the center-of-mass of the galaxy (either by construction of a simulation or by selection of a net galactic redshift in a survey), this $\velsym_{\mathrm{tot}}$ will implicitly contain the rotational velocity of the outflowing gas. This is particularly important at the base of the wind, where the gas still has all of its rotational motion. This is also important at large radii, where the rotation curve often flattens and the rotational velocity can be several hundred $\mathrm{km}~\mathrm{s}^{-1}$. Even more importantly, this can be the dominant component of $v_{\mathrm{tot}}$ if the wind is slow--the general prediction for CR-driven winds \citep[e.g.,][]{Peters2015_CRs_tallbox_xrays}. We therefore define an extra velocity $\bm{\velsym}_{i+\mathrm{rot}}\equiv\bm{\velsym}_{i}+(220~\mathrm{km}~\mathrm{s}^{-1}) \; \bm{e}_x$, where $220~\mathrm{km}~\mathrm{s}^{-1}$ is an approximate solar neighborhood rotational velocity, and the $x$-direction is chosen for simplicity. This velocity then goes into the definition for $\dot{E}_{\mathrm{out}}$ to produce the additional energy loading term shown in Figure~\ref{fig:height_avg_etas_and_sfr}. The ``Kin.+Rot.'' energy loading is always greater than the intrinsic kinetic energy loading, so we characterize this as a rotational ``boost.'' The ``Kin.+Rot.'' energy loading physically corresponds to the total kinetic energy of a parcel of gas launched from a source co-rotating with the galaxy.

For the points at single heights, we measure the fluxes through a slice, similar to \citet{Kim2020_SMAUG} and \citet{Rathjen2023}. This is numerically done by sampling the gas with a grid of points and replacing the cell-summations in Equation~\eqref{eq:mass_flux} and \eqref{eq:energy_flux} with this sampling, where total quantities divided by $\Delta z$ (the height of the volume-sampling) are replaced with densities times a small area element $\mathrm{d}A$ (e.g., $M_i/\Delta z \rightarrow \rho_i \mathrm{d}A$). We also include time-averaged data from Figure~\ref{fig:sfr_mdot_edot_time} to represent the average value of these quantities throughout the entire volume of the wind. Loading factors are first calculated per-snapshot, then averaged over time.

First looking at the SFR in Figure~\ref{fig:height_avg_etas_and_sfr} (upper), we find that the fluctuations in each of the cases can be consistent with each other, but the averages reflect the disparities. Both \crmhd cases are broadly consistent with the Kennicutt-Schmidt star formation rate for the initial surface density \citep{Kennicutt1998_KS_Law}. The \nlX case lies slightly closer to, but below, the Kennicutt-Schmidt relation. The \inX case is primarily above, and the \mhdX case is entirely above. However, we again caution against an absolute interpretation of these values due to the idealized nature of the setup.

The mass loading factors in Figure~\ref{fig:height_avg_etas_and_sfr} (center) show significant differences between the different simulations and different heights. The \mhdX case always produces the lowest mass loading factors, which decline significantly at $|z|=3~\mathrm{kpc}$. This is a result of the hot, thermally-driven wind scenario. The wind is primarily hot and low density, meaning the mass contribution to the mass flux is low. The wind is also not able to energize itself throughout its extent in the way that CRs can, meaning that the wind becomes weak by $|z|=3~\mathrm{kpc}$. This produces a very low average mass loading factor at this height. The two \crmhd cases, in contrast, show a less significant height-dependent reduction in mass loading. Between the two \crmhd cases, the \nlX case tends to have a higher mass loading. This is a consequence of both its lower SFR (decreasing the denominator in mass loading) and the better coupling of CRs to cold gas in the wind. As a result of fluctuations in the SFR and mass outflow rate shown in Figure~\ref{fig:sfr_mdot_edot_time}, the mass loading factors can occasionally exceed unity. All cases are, on average, consistent with a mass loading below unity, implying that the gas is primarily processed through the formation of stars. Only in the \nlX case does the mass loading temporarily deviate from this picture. The \nlX case produces a mass loading that is about 4.9 times that in the pure MHD case and 1.8 times that in the \inX case. The \inX scenario results in a mass loading about 2.7 times that seen in the pure MHD case.

In the bottom panel of Figure~\ref{fig:height_avg_etas_and_sfr}, we again see the specific behaviors of the kinetic and thermal energy fluxes seen in Figure~\ref{fig:sfr_mdot_edot_time}. Namely, in the \mhdX case, the wind is primarily thermal; in \nlX, the wind is primarily kinetic; and in \inX, the wind is intermediately thermal and kinetic, but slightly kinetic. We also see that the CR energy loading factors differ between the two \crmhd cases. The CR energy loading for the \inX case is, on average, slightly higher than that for the \nlX case. Because both of these energy loading factors are normalized by their respective SFRs, this means that the outflow in the \inX case carries a greater fraction of the injected CR energy than the fraction of injected CR energy carried in the \nlX outflow. This can be understood to be a result of CR loss terms. IND in the \inX case allows CRs to decouple from gas, reducing the pressure gradient and subsequent Alfv\'en wave losses. Additionally, the gas in the \nlX case is slightly denser on average than in the \inX case. This results in an increase of CR loss terms (hadronic, Coulomb, ionization) in the \nlX, which will also reduce the CR energy loading factor. Although both cases produce a CR energy loading very close to the injection efficiency (shown as a dashed line in Figure~\ref{fig:height_avg_etas_and_sfr}), the \inX case is closer. This also means that more of the CR energy can be transferred to higher altitudes, where global-scale wind-launching occurs as seen in \tttf. This is discussed in more detail in Section~\ref{sec:discuss_t24}.

We also see that the ``boosted'' energy loading factors, which additionally account for the rotational motions of gas near the solar circle, are much higher than the other thermal and kinetic energy loading values. This highlights the importance of making a distinction in how the kinetic energy flux is measured. If the rest frame velocity of the gas is taken to be the same for the entire galaxy, then the measured kinetic energy loading can become very large. If the rest-frame velocity is taken locally at the point of the outflow, this effect disappears. In this case, these values of the kinetic energy loading factors represent an upper limit given by an edge-on spectroscopic measurement. Notably, these ``boosted'' energy loading factors are more consistent with observations \citep{Xu2022_CLASSY_mass_loading} in the \nlX  and \inX  cases than in the pure MHD case. The ``boost'' resulting from the addition of rotational velocity can be approximated by taking a differential $\delta E_{\mathrm{kin,}\,i} = \delta(\frac{1}{2} \rho v_i^2) = \rho_i v_i \delta v_i$ at a constant density, where $v_i$ is the velocity in the co-rotating frame (the total velocity in our tallbox setup) and $\delta v_i$ is the additional rotational component of the motion. This approximation is justified by the fact that, typically, the outflow velocity near the base of the wind ($|z|\lesssim10~\mathrm{kpc}$) is much less than the rotational velocity \tttfp. Dense winds, that are present in the \nlX  and \inX  cases, tend to appear better augmented by the addition of the rotational motion.

\subsection{Outflow content}
\label{sec:outflow_content}

\begin{figure*}
\includegraphics{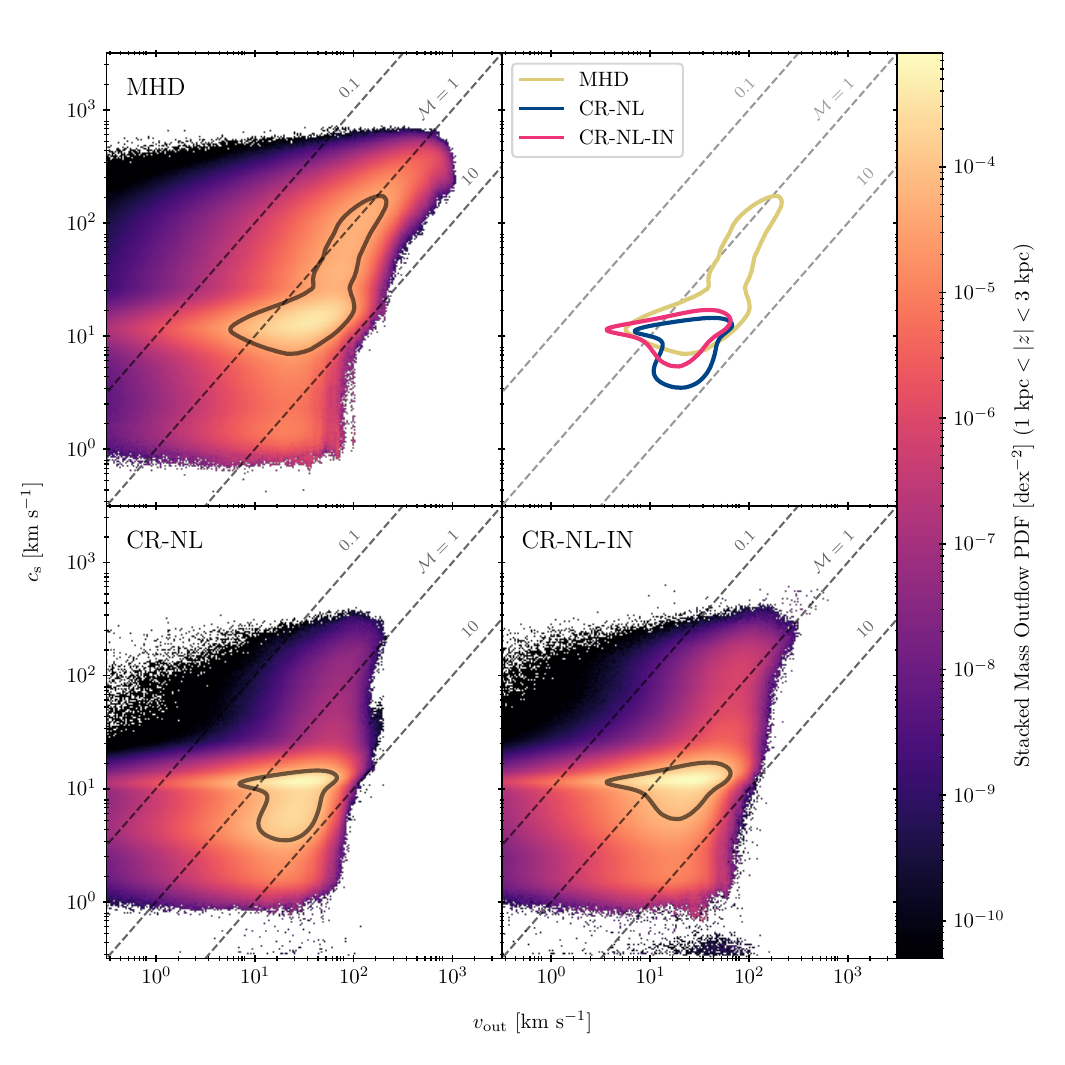}
\caption{Phase structure of outflowing material in the pure MHD case (upper left), \nlX (lower left), and \inX (lower right), binned by thermal sound speed and outward velocity. The color represents the PDF of outward mass flux measured over $112~\mathrm{Myr}<t<250~\mathrm{Myr}$. Dashed lines denote Mach numbers of 0.1, 1, and 10. The regions enclosed by the contours contain $75\%$ of the respective total mass fluxes, and are re-plotted together in the upper right panel. In the \mhdX case, the warm and hot phases are the main contributors to the mass flux, whereas the mass outflow in the \nlX and \inX cases occurs predominantly through the warm and cool phases.}
\label{fig:outflow_phase_pdf}
\end{figure*}

Figure~\ref{fig:outflow_phase_pdf} shows the distribution of outflowing material within $1~\mathrm{kpc}<|z|<3~\mathrm{kpc}$ binned by its outflow velocity $v_\mathrm{out}$ and its thermal (adiabatic) sound speed $c_\mathrm{s}$. The diagonal lines of slope 1 on this plot correspond to lines of constant Mach number. This is similar in construction to plots in \citet{Kim2020_TIGRESS_outflows}, \citet{Rathjen2023}, and \tttf. In all three star-forming cases, the outflow is primarily supersonic. The differences in the phase of the wind between the cases discussed in Section~\ref{sec:structure_and_profiles} are reflected clearly in the distributions shown here. The mass outflow in the pure MHD case is driven by the fast wind dominated by the warm/hot phase. Compared to other cases, the wind in the MHD case tends to be characterized by lower Mach numbers, i.e., the outflow is primarily dominated by thermal rather than kinetic energy.

The three star-forming cases all contain significant mass around $c_\mathrm{s}\sim10~\mathrm{km}~\mathrm{s}^{-1}$ (equivalent to  $T\sim10^{4}~\mathrm{K}$). However, the remainder of the mass flux differs between the physics cases. The \mhdX case carries mass outward via the warm and hot phases, but not the cold phase. Neither the \nlX nor the \inX case carry significant mass outward via the hot phase. This is both a consequence of their reduced SFR (meaning less hot gas from SNe) and the difficulty of CRs in accelerating a hot outflow \citep{Armillotta2024}. The \crmhd cases are instead able to drive mass outflow via the cool phase, particularly in the \nlX case. This general trend toward cooler CR outflows is seen in tallbox simulations utilizing purely diffusive CRs \citep{Rathjen2023} and global simulations \tttfp. We discuss the acceleration of gas by CRs in these outflows in Section~\ref{sec:outflow_driving}.

Comparing our pure MHD case to that in \citet{Rathjen2023} for the same initial surface density, we produce cold gas in the outflow that is not found in \citet{Rathjen2023}. This difference is likely attributed to both our radiation treatment and the geometric setup of the box. We do not explicitly perform radiative transfer for the FUV band and instead assume attenuation based on the initial conditions (Appendix~\ref{sec:fuv_appendix}). The \citet{Rathjen2023} \textsc{silcc} model includes tree-based radiative transfer for this band, meaning that optically-thin gas in the hot wind can permit FUV radiation to pervade the wind and evaporate cold gas. Our FUV is pre-attenuated, so this process cannot happen self-consistently. Additionally, our box is twice as wide in the $x-$ and $y-$directions (we use $1~\mathrm{kpc}$, while the \textsc{silcc} model assumes $512~\mathrm{pc}$). This increase in the box width permits the production of a horizontally-variable wind, resulting in the face-on spatially-variable distribution seen in Figure~\ref{fig:dens_projections}. This effect can also be seen when comparing TIGRESS solar-neighborhood simulations \citep[e.g.,][]{KimOstriker2017} with \textsc{silcc} simulations \citep[e.g.,][]{Rathjen2023}. In Figure~\ref{fig:dens_projections}, we see that the MHD case has an excavated region and two dense ``arms'' in the wind, whereas the gas distribution is more uniform in the horizontal direction in \citet{Rathjen2023}. Phase space distributions shown in Figure~\ref{fig:outflow_phase_pdf} implicitly sample a structurally different part of the wind compared to the \textsc{silcc} model. Additionally, simulations with CRs in \citet{Rathjen2023} have a higher average star formation than simulations without CRs. We instead find that our \mhdX simulation has the highest SFR, and that including CRs reduces the SFR. An increase in the SFR is theorized to assist in the survival of cold gas in a wind when the radiation field is appropriately scaled \citep{Vijayan2024_molecular_winds_theory}. Because our radiation field is static, this cold gas survival behavior could be further enhanced in the pure MHD case, allowing for a cool outflow that would not be found with the relative (``inverted'') SFRs from \textsc{silcc} or a self-consistent FUV field.

Compared to \tttf, outflow properties in our simulations are more diverse in all physics cases. In \tttf, no numerically-stable outflowing material with $c_\mathrm{s}<10~\mathrm{km}~\mathrm{s}^{-1}$ is present, whereas in our simulations low-temperature gas with $c_\mathrm{s}\sim1~\mathrm{km}~\mathrm{s}^{-1}$ can be found. This is an expected result of our increased resolution, as we are naturally able to retain smaller, denser clouds in the outflow. We also elect to measure the outflow at the base ($1~\mathrm{kpc}<|z|<3~\mathrm{kpc}$), which is representative of initial outflow-launching and consistent with similar work \citep[e.g.,][]{Rathjen2023}. The inner wind ($|z|\lesssim3~\mathrm{kpc}$) is denser and more fountain-like than the global-scale wind ($|z|\sim10~\mathrm{kpc}$; \tttf), so we expect to find colder, denser gas in our simulated winds compared to global-scale simulated galactic winds. In \tttf, the nominal numerical resolution reached above the galactic disk is enforced by a super-Lagrangian refinement technique to be close to $100~\mathrm{pc}$, whereas in our setup the entrained cold phase is resolved by cells with resolution down to $3~\mathrm{pc}$. The cold clouds seen in Figure~\ref{fig:dens_temperature_slices} are resolved by a few cells and hence are a few parsec in diameter, meaning that contemporary simulations of global-scale galaxies still lack the required resolution to faithfully capture these cold clouds and thus important aspects of the multiphase nature of galactic outflows. This highlights the necessity for performing detailed, highly resolved simulations of galactic environments, such as ours, to fill this gap. 

\begin{figure}
\includegraphics[width=\linewidth]{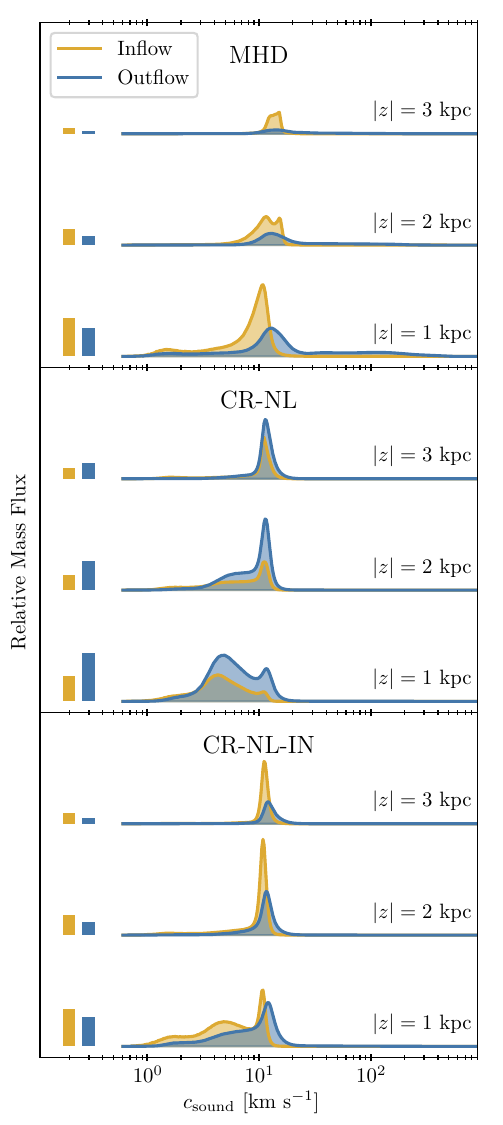}
\caption{Distribution of inflowing (yellow) and outflowing (blue) mass flux at different heights for pure MHD (upper), \nlX (center), and \inX (lower). Data is binned by its thermal sound speed. The total area under each blue and yellow curve is proportional to the outward and inward mass flux at a given height, respectively, and is represented by vertical bars on the left. The data shown is averaged between $112~\mathrm{Myr}$ and $250~\mathrm{Myr}$. The panels exhibit a diverse set of galactic fountain properties.
\label{fig:outflow_height_fountain}}
\end{figure}

In Figure~\ref{fig:outflow_height_fountain}, we compare the inflowing and outflowing material at different heights. We show the distribution of inward and outward mass flux binned by its sound speed and normalized by the total mass flux at a given height. Data is averaged over $112~\mathrm{Myr}<t<250~\mathrm{Myr}$. This is effectively a gas velocity-integrated version of Figure~\ref{fig:outflow_phase_pdf}, but with the addition of inflow data and data at different heights. Each case exhibits physically significant differences and the distributions of the inflowing and outflowing material are not perfectly correlated. This is the signature of a fountain flow, where gas is both inflowing and outflowing in a steady state. In addition to the mass inflow and outflow shown in Figure~\ref{fig:outflow_height_fountain}, there is also mass formed in stars. Therefore, star formation can be considered an additional ``sink'' for gas mass. This additional sink resolves the apparent violation of mass conservation implied by having a greater inflow mass flux than outflow mass flux. Although there is no net inflow from the boundary, the gas present in the wind can fall into the midplane and fuel star formation.

The pure MHD case (upper panel in Figure~\ref{fig:outflow_height_fountain}) produces a primarily warm outflow (by mass) with some hot and cold material as well. The outflow becomes almost exclusively warm by the time it reaches  $2~\mathrm{kpc}$, and nearly disappears beyond $3~\mathrm{kpc}$. At each height, the total inflowing mass flux is greater than the outflowing flux. The inflow is primarily warm with a cold component at low heights, where the hot phase is missing. At $2~\mathrm{kpc}$ and $3~\mathrm{kpc}$, the inflowing gas is exclusively warm. This reflects the thermal SN-driving scenario in which hot, fast gas can entrain gas of lower temperatures, but eventually cools above the midplane and subsequently inflows. The inflow and outflow curves at the base both show a prominent peak near $c_\mathrm{sound}\sim10~\mathrm{km}~\mathrm{s}^{-1}$. The separation between the peaks in the two curves could be a signature of hydrogen ionization at $\sim$$10^{4}~\mathrm{K}$. Specifically, the gas in the outflow may consist of primarily ionized hydrogen, and the inflowing material is primarily neutral hydrogen.

The \nlX case (middle panel in Figure~\ref{fig:outflow_height_fountain}) is very dissimilar from both other cases. The outflow is primarily cool with a warm component. The cool component remains significant up to heights of $2~\mathrm{kpc}$, but is overshadowed by the warm component at larger distances from the disk midplane. The outflow is sustained up to $3~\mathrm{kpc}$ and is nearly exclusively warm. In this scenario, at every height, the mass flux in the outflow is greater than in the absolute value of mass flux in the inflow. This is the only case where the mass loading can be above unity, implying a greater fraction of baryons can be instantaneously processed in outflows than in star formation. An excess of inflowing material requires that this material be processed in stars in a steady state to prevent runaway accretion of gas. Because there is no excess of inflowing material, the star formation rate is reduced. This wind is primarily cold and CR-driven, and therefore entirely different from the pure MHD case. Cool and warm material is continuously accelerated throughout the wind and is able to flow out unimpeded.

The \inX case (lower panel in Figure~\ref{fig:outflow_height_fountain}) is more similar to the pure MHD case than \nlX case. The base of the outflow is both warm and cool, but becomes predominantly warm at higher heights. The absolute value of the mass flux in the inflow is always greater than the mass flux in the outflow, the inflowing gas tends to be colder, and the peaks in the mass flux distributions are offset with respect to each other near the sound speed corresponding to the gas temperature for which the hydrogen ionization rate peaks. There is no immediate explanation for this phase transition. It is feasible that the gas in the fountain flow undergoes a quasi--parabolic trajectory where the gas simultaneously cools and loses outward momentum. A single parcel of gas following this trajectory would produce a warmer peak in the outflow distribution and a colder peak in the inflow distribution. This process cannot definitively be attributed to IND, as the offset in the peaks of the mass distributions appears similar in the \mhdX and the \inX cases, implying it does not necessarily depend on CRs or IND. Another possibility is that this apparent transition is driven by the mixing of cold clouds with warm/hot gas. Cold clouds may be ``shredded'' in the outflow, subsequently mixing with the surrounding warm/hot gas and producing cooler gas that is no longer outflowing. The kinematics of the multiphase wind are discussed further in Section~\ref{sec:velocity_profiles}.

The magnitudes of both inflow and outflow mass flux at $3~\mathrm{kpc}$ are greater in the full-physics case when compared to the pure MHD case. The wind in the \inX case is CR-driven, allowing for the continued acceleration of warm gas up to the height of $3~\mathrm{kpc}$. However, the inclusion of IND makes acceleration of warm neutral gas more difficult than in the \nlX case. This difference is discussed in detail in the following section.

\section{Driving multiphase outflows}
\label{sec:outflow_driving}

In this section, we investigate the multiphase galactic winds. We compare the acceleration by CRs between the \nlX and \inX cases, and look at the resulting outflow velocities for all cases. We adopt phase bins from ~\citet{KimOstriker2017} for simplicity and intuitive comparison, although we acknowledge that a more sophisticated binning \citep[e.g.,][]{Kim2023_tigress_NCR} is more appropriate for a simulation with a non-equilibrium ionization model. The definitions of our adopted phase bins are included in Figure~\ref{fig:eddington_cr_phase}.

\subsection{CR acceleration of outflows}
\label{sec:cr_accel_gas}

\begin{figure}
    \includegraphics[width=\linewidth]{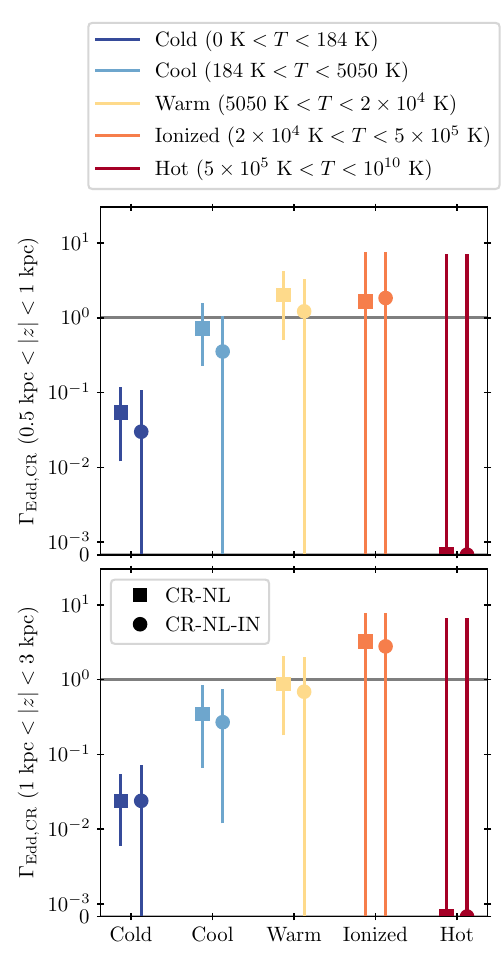}
    \caption{CR acceleration Eddington factors at the base of the wind ($0.5~\mathrm{kpc} < |z| < 1~\mathrm{kpc}$; upper) and the body of the wind ($1~\mathrm{kpc} < |z| < 3~\mathrm{kpc}$; lower), separated by phase. The CR acceleration Eddington factor is defined as $\Gamma_\mathrm{Edd,CR}=-a_{\mathrm{CR},z}/a_\mathrm{grav}$, where $a_{\mathrm{CR},z}$ and $a_\mathrm{grav}$ are the CR and gravitational acceleration, respectively. Eddington factor values above unity signify a net outward acceleration by CRs. Volume-weighted median and $25^{\mathrm{th}}$ and $75^{\mathrm{th}}$ percentiles of fluctuations over $112~\mathrm{Myr}<t<250~\mathrm{Myr}$ are displayed as points with error bars. CRs are better suited to accelerate warm and ionized gas, and supply no momentum to the hot phase. The cool and cold phases are lightly supported against freefall.
    \label{fig:eddington_cr_phase} }
\end{figure}

We focus our analysis on the acceleration provided by CRs. Although thermal and magnetic forces can contribute to the acceleration of outflows \tttfp, the objective of this work is to characterize the impact of CRs on a multiphase galactic wind in high-resolution simulations. The critical difference between the \nlX and \inX cases is the inclusion (or omission) of IND. So far, we have quantified how this affects the properties and content of the respective galactic winds. Now, we characterize the differences in acceleration of the gas by CRs.

Figure~\ref{fig:eddington_cr_phase} shows the volume-weighted CR Eddington factor and $25^{\mathrm{th}}$ to $75^{\mathrm{th}}$ percentiles of fluctuations in the \nlX and \inX runs, binned by phase. We analyze the base of the wind ($0.5~\mathrm{kpc} < |z| < 1~\mathrm{kpc}$; lower panel in Figure~\ref{fig:eddington_cr_phase}) and the body of the wind ($1~\mathrm{kpc} < |z| < 3~\mathrm{kpc}$; upper panel in Figure~\ref{fig:eddington_cr_phase}) with phase definitions from \citet{KimOstriker2017}. We define the Eddington factor corresponding to acceleration by CRs as $\Gamma_\mathrm{Edd,CR} \equiv -a_{\mathrm{CR},z}/a_\mathrm{grav}$, where $a_{\mathrm{CR},z}$ is the vertical acceleration of the gas by CRs, and $a_\mathrm{grav}$ is the gravitational acceleration from both self-gravity between active material in the simulation and the applied external potential from Equation~\eqref{eq:smoothed_grav_accel}. As in \tttf, we compute the vertical acceleration by CRs as $a_{\mathrm{CR},z}=(\boldsymbol{\nabla} P_\mathrm{CR}/\rho) \bm{\cdot} \bm{e}_z$.

The upper panel of Figure~\ref{fig:eddington_cr_phase} shows the average CR Eddington factor at the base of the wind ($0.5~\mathrm{kpc}<|z|<1~\mathrm{kpc}$) and reveals (i) significant differences between the acceleration of different phases, and (ii) differences between the \nlX and \inX cases. In the \nlX case, cold, cool, and warm gas are more efficiently accelerated than in the \inX case. Cold gas is not super-Eddington in the base of the wind in either case. The \nlX case is slightly better at accelerating cold gas than the \inX case--an outcome of including IND. The trends for the base of the wind are generally mirrored for the body of the wind (lower panel in Figure~\ref{fig:eddington_cr_phase}). Eddington factors for cold, cool, and warm gas are all lower in the body of the wind than at the base of the wind. This is likely due to CRs losing energy throughout the wind, meaning that there is less CR momentum available to be deposited into the gas. Interestingly, CRs lose more energy in the \nlX case in comparison to the \inX case because of the tight coupling of CRs to the surrounding plasma in this model, which amplifies CR Alfv\'en wave losses.

We note that CRs generally do not accelerate hot gas. This is in agreement with \citet{Armillotta2024}, who find that CRs tend not to accelerate a hot wind. \citet{Armillotta2024} find this to be the result of CR advection in the rapidly-moving hot gas. This can cause a net momentum transfer from the gas to CRs and a CR pressure gradient reversal, which then acts against the wind. We find this to be a plausible explanation for the vanishing Eddington factors observed in hot gas.

\subsection{Gas velocity by phase}
\label{sec:velocity_profiles}

\begin{figure}
\includegraphics{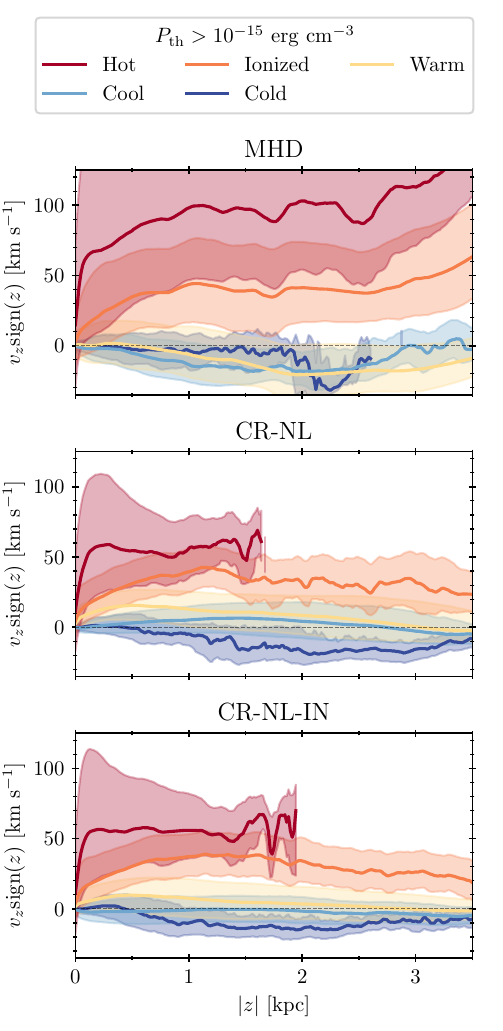}
\caption{Profiles of volume-weighted velocity separated by phase for the pure MHD case (top panel), \nlX (center pane), and \inX (bottom panel) cases. The bold lines represent median values over $112~\mathrm{Myr}<t<250~\mathrm{Myr}$, and the shaded region encloses fluctuations within the $25^{\mathrm{th}}$ and $75^{\mathrm{th}}$ percentiles. A pressure floor of $10^{-15}~\mathrm{erg}~\mathrm{cm}^{-3}$ and a minimum Voronoi cell count per bin of 150 is adopted to filter only physical results. The phase colors are repeated at the top of the figure, with temperature ranges given in Figure~\ref{fig:eddington_cr_phase}. The gas of different phases settles into distinct kinematic channels. The \nlX and \inX cases exhibit warm and cool phases with a net outward velocity higher than that of the case without CRs. 
\label{fig:velocity_phase_profiles}}
\end{figure}

In Figure~\ref{fig:velocity_phase_profiles}, we show the vertical velocity profiles of gas separated by phase. We display the volume-weighted median and $25^{\mathrm{th}}$ to $75^{\mathrm{th}}$ percentiles of fluctuations. The gas of different temperatures clearly settles into distinct steady-state kinematic behaviors.

Hot and ionized gas are the fastest phases in all cases. In the pure MHD case, the hot phase is faster than in both \nlX and \inX case. This is a consequence of the higher rate of SNe, and the rapid propagation of freshly-heated gas through the already hot low-density wind.
The ionized phase could either be cooled from the hot phase, or CR-heated from the warm phase. In the former case, the ionized phase should inherit the rapid velocity from the hot phase. In the latter case, the ionized gas would be accelerated by CRs to a higher velocity than the warm phase. In Figure~\ref{fig:velocity_phase_profiles}, we see that the ionized phase primarily has a velocity between the warm and the hot phases. This is in agreement with the cooling scenario and the CR-heating scenario.

The velocity profile of the warm phase is significantly different between the case without CRs and either the \nlX or \inX case. In the pure MHD case, the warm phase is preferentially inflowing with a slight trend to zero velocity toward the midplane. The \nlX and \inX cases, instead, exhibit a preferentially outflowing warm phase out to $\sim$$3~\mathrm{kpc}$. This is due to the CR force that is counteracting gravity. The Eddington factor for the CR acceleration of the warm phase is preferentially slightly below unity in the body of the wind (lower panel in Figure~\ref{fig:eddington_cr_phase}), so the warm gas slowly loses velocity as it travels outward.

The cool phase is also preferentially inflowing in the case without CRs, but is slightly different between the \nlX and \inX simulations. In the \nlX case, the cool phase is preferentially traveling outward up to $~\sim$$2.6~\mathrm{kpc}$, beyond which it shows a slight tendency for inflow. In the \inX case, the velocity of the cool phase hovers near and slightly below zero. As shown in Figure~\ref{fig:eddington_cr_phase}, in the \nlX case, while cool gas is super-Eddington at the base of the wind, it can accrete from larger heights. When IND is included, the cool phase is instead levitated throughout the base and body of the wind.

The cold phase in the pure MHD case (Figure~\ref{fig:velocity_phase_profiles}; upper panel) has preferentially zero velocity, likely due to a combination of magnetic effects (tension, draping) and cloud-crushing physics within the hot wind. The negative portion of the velocity of the cold phase at $\sim$$2~\mathrm{kpc}$ could be a consequence of a fountain flow or in-situ formation of molecular gas via thermal instability. In the \nlX and \inX cases, cold gas is preferentially inflowing at the body of the wind as the CR pressure gradient is not effective at accelerating the cold phase.

Comparing the velocities of the warm and ionized phases, we see that the ionized phase always has a higher outflow velocity than the warm phase. The ionized phase is not inflowing, whereas the warm phase often has an inward velocity. Together, this implies a scenario where hotter, ionized gas is participating in a net outflow while warm gas can simultaneously flow inwards. This is in agreement with the quasi-parabolic trajectory of gas proposed in Section~\ref{sec:outflow_content}. However, there is also evidence for the mixing scenario. If the outflowing ionized phase were to mix with low-velocity cool/cold clouds, the resulting gas would have an intermediate temperature and small outflow velocity. It is still not apparent whether the quasi-parabolic trajectory or mixing scenario, or both simultaneously explain the mass fluxes in Figure~\ref{fig:outflow_height_fountain}.

This data is partially sensitive to numerical effects. The fountain flow in the pure MHD case would produce a net mass inflow from the boundary if inflow from the boundary was allowed. However, as we enforce outflow-only boundaries, this net inflow does not occur. Therefore, when the gas is subject to net inward acceleration, it can become rarefied near the boundary, producing a cold, vacuum-like region. Because the inflowing gas is low-density, it does not affect any mass-weighted quantities, and does not significantly impact the kinematics of the outflow. However, the low-density gas is volume-filling and becomes relevant for volume-weighted velocity profiles. To filter out this cold, low-density gas, we adopt a thermal pressure floor of $10^{-15}~\mathrm{erg}~\mathrm{cm}^{-3}$ for Figure~\ref{fig:velocity_phase_profiles}. We also require that each bin along the $z$-axis contain 150 unique Voronoi cells over the full time analysis to be displayed, which filters out small-number transient hot outflow events. We confirm that the velocity profiles are robust to these data-cleaning prescriptions.

\section{Comparison with Global Simulations}

\label{sec:discuss_t24}
\indent
We now compare the results from our tallbox simulation suite to those presented in \tttf, who perform global simulations of galactic winds. Unlike in \tttf, due to significantly enhanced numerical resolution adopted in our work, our simulations exhibit a multiphase wind (see Figure~\ref{fig:dens_temperature_slices}) and resolve both the dense molecular clouds and hot SNRs in the midplane. Our higher resolution is better suited to tracking cosmic ray transport and confinement within the interstellar medium and the inner wind. In particular, we are able to capture CR interactions with very cold gas, which is known to be important for CR losses in dense media \citep{BustardZweibel2021}. Our resolution also enables us to capture aspects of CR self-confinement in tenuous, shock-heated gas, expected to be lacking in global-scale simulations as described in \citet{Armillotta2024}. Specifically, the energy-driven stage of SNRs with very low densities create regions where CR transport is NLLD-dominated, but is still described by very low diffusion coefficients \citep{Armillotta2021}.

However, our approach is not suited for predicting global scale magnetic field topology, which can be important in determining CR transport regimes \citep{Armillotta2021,ThomasPfrommerPakmor2023_Wave_Dark}. The geometry of the tallbox approach additionally precludes the study of global consolidation of an outward-propagating wind and spherical outflow expansion through a sonic point \citep{RuszkowskiPfrommer2023_CR_Review}. Therefore, ours is not a flawless extension of earlier work to higher resolution, but rather a look at the small-scale processes that are currently unresolved in global simulations.

We differ slightly from \tttf in our result for the feedback regulated SFR between the \mhdX and \inX case. \tttf find a very slight difference between the respective SFRs, which does not produce a visible disparity between the face-on projections. Their interpretation is that CR feedback has only a minor impact on the global radial structure of the ISM and, in particular, it does not change the occurrence rates and visual indicators of stellar activity including superbubble sizes. We instead find that the difference between the feedback activity in respective steady states produces differences in face-on projections. In particular, in the \mhdX case, SNe can very efficiently displace ISM producing large superbubbles. This is likely a consequence of the fact that we analyze our simulations at earlier times than \tttf~($250~\mathrm{Myr}$ considered here as compared to $1~\mathrm{Gyr}$ in \tttf). Our SFRs are different on average in the different physics cases. The SFRs in \tttf~converge to a single value, with and without CRs. It is feasible that our SFRs would converge to a single value after an additional $\sim750~\mathrm{Myr}$ of evolution. We also note that the size of superbubbles, both simulated \tttfp~and observed \citep{Barnes2023}, can be comparable to the horizontal extent of our simulation box. Therefore, the tallbox setup precludes morphological study of the superbubbles.

\tttf also find that the wind undergoes a significant transition by $|z|\sim10~\mathrm{kpc}$, which is far beyond the height of our tallbox. At this height, the global galactic wind is accelerated in both of the simulations presented in \tttf. The important connection between the small-scale tallbox wind and the global-scale galactic wind is that the tallbox wind at $4~\mathrm{kpc}$ (the vertical distance between the midplane and the edge of the box) must be able to supply more than the total amount of energy flux at $10~\mathrm{kpc}$, as energy will only be dissipated while the wind material travels from a height of $4~\mathrm{kpc}$ to $10~\mathrm{kpc}$.

We find that, in the \nlX and \inX cases, the energy in the winds is carried away primarily by CRs rather than in the thermal or kinetic form. These CR-driven winds can sustain significant mass loading factors out to $3~\mathrm{kpc}$. In the \mhdX case, mass and energy fluxes are much smaller at the height of $3~\mathrm{kpc}$ compared to those present in the \nlX and \inX cases. If we consider all of the mass and energy to be perfectly transported out to $10~\mathrm{kpc}$ (as an upper limit for global wind-driving), the pure MHD case would be unable to launch a large-scale galactic wind. In the \nlX and \inX cases, however, considerable amounts of energy can be transported to that height, which can then be expended to accelerate the gas as in \tttf. Outflows can carry significant a fraction of energy in the CR form, especially in the \inX case. Our results reflect a failure of the thermally SN-driven wind paradigm under our assumptions. CR-driven winds, in contrast, have no trouble supplying the material to the global-scale wind.

\section{Summary and Conclusions}
\label{sec:summary}
We present the first self-consistent and feedback-regulated tallbox simulations with two-moment CR transport model including different wave damping mechanisms and CRs dynamically coupled to the gas. We use the moving mesh MHD code \textsc{Arepo} with the \textsc{Crisp} feedback model. We quantify the impact of sequential additions of CR microphysics, namely purely advective CRs, two-moment CR transport with only NLLD, and full-physics CR transport with both NLLD and IND. We showcase the complicated effects of feedback in a multiphase ISM and the resulting galactic winds. An itemized summation of our findings is as follows.

\begin{enumerate}

    \item The \mhdX case, \nlX case, and the \inX case all result in a star-forming ISM with a galactic wind. IND does not completely suppress the efficacy of CR feedback. Despite the fact that IND effectively decouples CRs from neutral gas, CRs with IND are able to drive a warm galactic wind with moderate mass, kinetic energy, and thermal energy loading factors.
    
    \item CRs reduce the SFR by decreasing the amount of gas available to form stars. This happens because CR pressure replaces thermal pressure and prevents gas collapse into molecular clouds. When IND is included, CRs decouple from the cold regions allowing the gas to condense, cool, and form stars. Despite the fact that IND is also included in global simulations, the impact of IND on star formation is not captured in global simulations, thus underscoring the need for adopting adequately high numerical resolution to capture these trends.
    
    \item Each physics case shows evidence of a galactic fountain, where gas is simultaneously inflowing and outflowing. The mass and energy loading factors tend to be smaller at $|z|=3~\mathrm{kpc}$ than $|z|=1~\mathrm{kpc}$. In the \mhdX case, this vertical drop-off is much more significant than in the \nlX and \inX cases. The \nlX case is the only star-forming case in which the mass outflow rate can exceed the SFR, albeit only temporarily.
    
    \item The additional velocity component from galactic rotation can be a significant component of the measured kinetic energy loading factor. It is critically important to consider this effect when calculating energy loading based on a single reference frame for the entire galaxy. When implementing this boost into our tallbox results, we find that the kinetic energy loading factors can be boosted by several orders of magnitude, especially in the dense winds produced in the \nlX and \inX cases. This effect must therefore be carefully accounted for when connecting local-scale observations (i.e., observations of a sub-section of a galaxy) and simulations (i.e., zoom-in simulations), in which the rotational component of the gas motion may be inadvertently ``absorbed" into the recessional velocity of the galaxy, with global-scale calculations that self-consistently model the effect of rotation.
    
    \item In the \nlX and \inX cases, energy is primarily transported outward by CRs, as indicated by both the dominance of CR pressure at all heights and the CR energy loading factors. A higher fraction of CR energy is carried out in the \inX case than in \nlX case. We attribute this, in part, to the tight coupling of CRs to the gas in the \nlX case, which leads to steeper CR pressure gradients and causes greater CR streaming losses than in the \inX case. Additionally, the increased gas density in the \nlX case amplifies hadronic, Coulomb, and ionization CR loss terms, which also contributes to the reduction in CR energy loading. The high outflow fraction of CR energy implies that a substantial portion of the CR energy generated in the disk is available for driving winds at greater altitudes \tttfp~and for pressurizing the CGM with CRs.
    
    \item CRs most effectively accelerate warm and ionized gas in outflows. We find that CRs do not assist in accelerating a hot outflow, in agreement with \citet{Armillotta2024}. CRs can also provide some momentum to cold and cool gas, but this effect is reduced by the inclusion of IND. CRs with IND, however, can still levitate cold gas in the wind by counteracting gravity. The acceleration of cold gas in the pure MHD case is likely related to the higher SFR and ability to entrain molecular gas within this hot wind. The survival of the cold gas in the wind in the \mhdX scenario is also likely a consequence of our simplified radiation treatment, where the escape of dissociating radiation in hot, optically thin gas is not properly modeled.
    
\end{enumerate}
\indent
An unexplored component of this work is the modes and mechanisms of CR transport, including effective diffusion coefficients. We have elected to present this work as a self-contained analysis of outflows, and reserve the characterization of CR transport for future work. We have shown that self-consistent simulations with CR feedback produce a galactic wind that is distinct from pure MHD simulations. Therefore, scattering coefficients by phase inferred from post-processed simulations in \citet{Armillotta2021} may need to be re-evaluated. We also neglect to compare our results with direct observables such as absorption and emission lines. Comparison with observations will be critical for confirmation of this work. Improvements on this work will require self-consistent radiation treatment and shearing box terms \citep[e.g.,][]{Kim2023_tigress_NCR} to comprehensively model the wind-driving environment.

In summary, the different physical models exhibit critical differences in feedback, which alter star formation rates and galactic outflows. CR feedback is well-suited to producing galactic outflows and can better supply energy to a global-scale wind than a purely thermally-driven wind. Importantly, the inclusion of IND does not result in complete suppression of the efficacy of CR feedback.

\section*{Acknowledgments}

MR thanks Lucia Armillotta, Norm Murray, Peng Oh, Eve Ostriker, Todd Thompson, and Ellen Zweibel for stimulating discussions.

This work was performed in part at the Kavli Institute for Theoretical Physics (KITP) during the ``{\it Turbulence in Astrophysical Environments}" program. This research was supported in part by grant NSF PHY-2309135 to the KITP.

This work was performed in part at the Aspen Center for Physics (ACP) during the ``{\it Cosmic Ray Feedback in Galaxies and Galaxy Clusters}" summer program. The ACP is supported by the National Science Foundation grant PHY-2210452, and by grants from the Simons Foundation (1161654, Troyer) and Alfred P. Sloan Foundation (G-2024-22395). 

MR acknowledges the support from the Leinweber Center for Theoretical Physics, which sponsored the organization of the ``{\it 7th ICM Theory and Computation Workshop}" during which BS, MR, and CP were able to discuss key aspects of the project.

MR acknowledges support from the National Aeronautics and Space Administration grant ATP 80NSSC23K0014 and the National Science Foundation Collaborative Research Grant NSF AST-2009227. TT, CP, and MW acknowledge support from the European Research Council under ERC-AdG grant PICOGAL-101019746. This work was supported by the North-German Supercomputing Alliance (HLRN) under project bbp00070.

This work used the Delta system at the National Center for Supercomputing Applications through allocation PHY240111 from the Advanced Cyberinfrastructure Coordination Ecosystem: Services \& Support (ACCESS) program, which is supported by National Science Foundation grants \#2138259, \#2138286, \#2138307, \#2137603, and \#2138296 \citep{ACCESS}. This research was also supported in part through computational resources and services provided by Advanced Research Computing (ARC)--a division of Information and Technology Services (ITS) at the University of Michigan, Ann Arbor.

\appendix

\section{Attenuated FUV Field}
\label{sec:fuv_appendix}

Radiation in the FUV band is an important regulator of the ISM thermochemistry. It is responsible for gas heating by the photoelectric effect on dust grains but also influences the chemical composition of the ISM by dissociating molecules and ionizing atoms with low ionization potentials such as C and Si. We account for FUV radiation in an idealized manner. We assume that FUV radiation is radiated by stars close to the midplane, which is subsequently absorbed by dust using an absorption cross-section of $\sigma_\mathrm{dust} = 1.57 \times 10^{-21}~\mathrm{cm}^{2}~\mathrm{H}^{-1}$. We construct our general solution of the radiative transfer equation using the relation between the energy density at a point of interest emitted by a source with luminosity $L$:
\begin{equation}
    u = \frac{L}{4 \pi s^2 c} \exp{(-\tau)} \label{eq:radtrans_single_source},
\end{equation}
where $s$ is the distance between the point of interest and the source while $\tau$ is the optical depth. We calculate the strength of the FUV radiation field in the limit of a source distributed over an infinitely thin surface located on the midplane and assume that the absorbing gas is distributed following the density distribution of our initial conditions. In this configuration, there is no single source but infinitely many sources located at the midplane. In this case, Equation~\eqref{eq:radtrans_single_source} becomes 
\begin{equation}
    u = \int \mathrm{d} A ~ \frac{(L/A)}{4 \pi s^2 c} \exp{(-\tau)}, \label{eq:radtrans_plane_source}
\end{equation}
where $(L/A)$ is the surface emissivity and the integral ranges over the entire midplane. Accounting for the rotational-translational symmetry of the midplane, the solution for the energy density $u$ of the radiative transfer equation is only dependent on the vertical height $z$ and reads
\begin{equation}
    u(z) = \frac{(L/A)}{2 c} \int_0^\infty \mathrm{d} R R ~ \frac{1}{R^2 + z^2} \exp{(-\tau)},
    \label{eq:fuv_final_integral}
\end{equation} 
where $s = \sqrt{R^2 + z^2}$, the area element is given by $\mathrm{d} A = 2\pi R\mathrm{d}R$, and the optical depth can be calculated from
\begin{equation}
    \tau = \int_0^{\sqrt{R^2 + z^2}} \mathrm{d}s ~\sigma_\mathrm{dust} n_\mathrm{H} \left(\frac{s}{\sqrt{R^2 + z^2}} \right).
\end{equation}

\begin{figure}
    \vspace{4mm}
    \includegraphics[width=\linewidth]{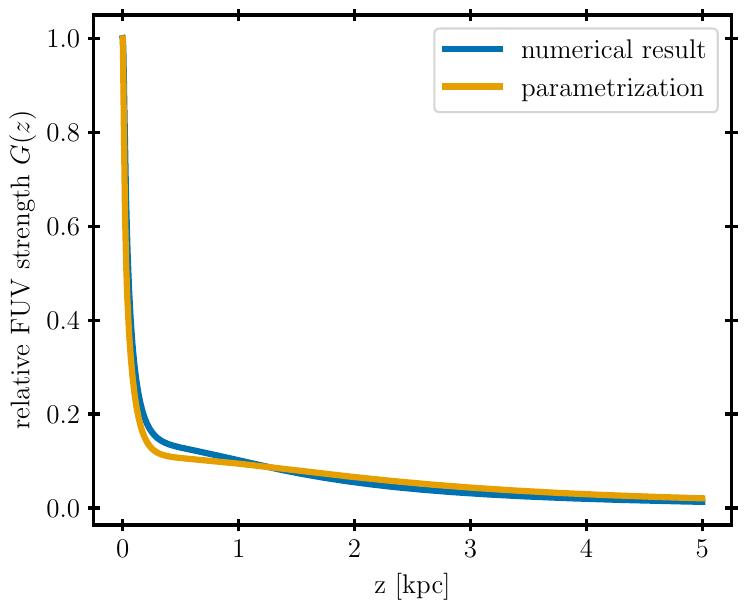}
    \caption{Comparison between the numerically integrated relative FUV radiation field strength (blue line) as defined by Eq.~\eqref{eq:relative_fuv_strength} and our three-component parametrization (orange line) as given by Eq.~\eqref{eq:G_components_sum}. }
    \label{fig:fuv_attenuation_fit}
\end{figure}

Formally, the integral for the energy density in Equation~\eqref{eq:fuv_final_integral} diverges logarithmically for $z \to 0$. We avoid this integral by imposing a physical value for $u(0)$ and renormalizing $u(z)$ by this value. To realize the envisioned solar-neighbourhood conditions in our setup, we assume that the unattenuated FUV radiation field close the midplane is described by the Habing field \citep{Habing1968_FUV_Normalization}  and we normalize our results by the energy density of the Habing field $u_\mathrm{Habing}$ using
\begin{equation}
u_\mathrm{Habing} = u(0) = \frac{(L/A)}{2 c} \int_0^\infty \mathrm{d} R R ~ \frac{1}{R^2} \exp{(-\tau)}
\end{equation}
and define $G$ to be the relative field strength of the energy density with respect to the midplane and the Habing field:
\begin{equation}
    G(z) =~\frac{u(z)}{u_\mathrm{Habing}} = \frac{\textstyle \int_0^\infty \mathrm{d}R R ~ \frac{1}{R^2 + z^2} \exp(-\tau)}{\textstyle \int_0^\infty \mathrm{d}R R ~ \frac{1}{R^2} \exp(-\tau)}, \label{eq:relative_fuv_strength}
\end{equation}

The hydrogen number density of the initial conditions is given by
\begin{equation}
    n_\mathrm{H}(z) = X \rho(z) / m_\mathrm{p},
\end{equation}
where $X = 0.76$ is the hydrogen mass fraction, and $m_\mathrm{p}$ is the proton mass, and $\rho(z)$ is given in Equation~\eqref{eq:sech_profile}. The optical depth evaluates to
\begin{equation}
    \tau = \sigma_\mathrm{dust} X \frac{\Sigma_\mathrm{gas}}{2} \frac{\sqrt{R^2 + z^2}}{z} \tanh\left(\frac{z}{2 z_0}\right),
\end{equation}
where we neglected the factor $\coth(H / 4 z_0)$ from the original density equation because it evaluates to $\sim 1 + 4\times 10^{-9}$ for our parameters $H=4~\mathrm{kpc}$ and $z_0=100~\mathrm{pc}$. We calculate the remaining integral defining $G(z)$ numerically but only over a range up to $R = 2$ kpc. This restricted integration domain is chosen to counteract the unphysical number of infinite copies of each source that we would need to account for in our periodic tallbox setup. In actual galaxies, the radial galactic length scale sets a typical scale for which the number of sources, and hence the strength of the diffuse FUV field (away from direct sources), changes. Furthermore, dust absorption gives rise to another relevant length scale -- under typical ISM conditions ($n_\mathrm{H} \sim 1~\mathrm{cm}^{-3}$) an optical depth of $\tau \sim 1$ is reached on kpc-scales and hence the contribution of sources beyond this scale is heavily subjected to absorption loses, which we cannot reliably account for in our simplified calculation that is solely based on the initial gas distribution. The logarithmic divergence at the lower integration bound is avoided by starting the integration at $R = 10~\mathrm{pc}$. Sources closer than this distance can be regarded as close-by stars and hence point sources which we do not explicitly account for in our modelling. Properly following the radiation from these nearby stars would require a special treatment outside of our model for the diffuse FUV background field. We parameterize the numerical result using three components 
\begin{subequations}
\begin{align} 
  \label{eq:G_component_a}
  G_a(z) &= \frac{G_1}{1 + (|z|/h_1)^{\alpha_1}},\\
  \label{eq:G_component_b}
  G_b(z) &= \frac{G_2}{1 + (|z|/h_2)^{\alpha_2}},\\
  \label{eq:G_component_c}
  G_c(z) &= (1 - G_1 - G_2) \exp{(-|z| / h_0)},\\
  \label{eq:G_components_sum}
  G(z) &= G_a(z) + G_b(z) + G_c(z), 
\end{align}
\end{subequations}
where the $G_c$ component matches the low-altitude exponentially-declining contribution of $G(z)$, and the two other components $G_a$ and $G_b$ approximate the high-altitude parts of $G(z)$. We use the following parameters 

\begin{align}
\begin{array}{lll}
    h_0 = 743  ~\mathrm{pc},  & & \\
    h_1 = 14.5 ~\mathrm{pc}  & \alpha_1 = 1.98 & G_1 = 0.397, \\
    h_2 = 2.44 ~\mathrm{kpc} & \alpha_2 = 2.04 & G_2 = 0.109, \\ \ 
\end{array}
\end{align}
and compare in Fig.~\ref{fig:fuv_attenuation_fit} the numerical result with our parameterization.

\bibliography{main}{}
\bibliographystyle{aasjournal}

\end{document}